\documentclass[fleqn,usenatbib]{mnras}

\usepackage{newtxtext,newtxmath}

\usepackage[T1]{fontenc}
\usepackage{ae,aecompl}

\usepackage{graphicx}	
\usepackage{amsmath}	
\usepackage{amssymb}	

\title[Gravitational recoil in E1821+643]{The spatially offset quasar E1821+643: New evidence for gravitational recoil}

\author[Yashashree Jadhav et al.]{
Yashashree Jadhav,$^{1}$\thanks{E-mail: ysj1195@rit.edu}
Andrew Robinson,$^{2}$
Triana Almeyda$^{3}$
Rachel Curran$^{2}$
\newauthor
and Alessandro Marconi$^{4}$
\\
$^{1}$Astronomy Program, Department of Physics and Astronomy, Seoul National University, 1 Gwanak-ro, Gwanak-gu, Seoul 08826, Korea\\
$^{2}$School of Physics and Astronomy, Rochester Institute of Technology, 1 Lomb Memorial Drive, Rochester, NY 14623, USA\\
$^{3}$Department of Biological and Physical Sciences, South Carolina State University, Orangeburg, SC 29117,USA\\
$^{4}$University of Florence, Firenze, Florence
}
\date{Accepted XXX. Received YYY; in original form ZZZ}

\pubyear{2020}

\begin{document}
\label{firstpage}
\pagerange{\pageref{firstpage}--\pageref{lastpage}}
\maketitle

\begin{abstract}
A galaxy merger is expected to cause the formation of a supermassive black hole (SMBH) binary, which itself eventually coalesces through the anisotropic emission of gravitational waves. This may result in the merged SMBH receiving a recoil kick velocity $\sim 100 - 1000$\,kms$^{-1}$, causing it to oscillate in the gravitational potential of the host galaxy.
The luminous quasar E1821+643, identified as an SMBH recoil candidate via spectropolarimetry observations, shows Doppler shifting of the broad emission lines in direct and scattered light, consistent with a relative velocity of 2100\,km\,s$^{-1}$ between the quasar nucleus and host galaxy. In this paper, we attempt to detect the expected spatial displacement using a combination of optical spectroastrometry and Hubble Space Telescope (HST) narrow band images. The spectroastrometry reveals a relative spatial displacement between the quasar nucleus and the gas emitting the [OIII]$\lambda\lambda 4959,5007$ lines of $\sim 130$mas ($\sim 580$\,pc) to the North-West. Our HST images resolve the [OIII] emission on sub-arcsecond scales, showing that it is asymmetrically distributed, extending to radial distances $\sim 0.5-0.6$\arcsec  from the nucleus in a wide arc running from the North-East around to the West. A simulated spectroastrometry observation based on the HST [OIII] image indicates that only a small fraction of the measured displacement can be attributed to the asymmetric [OIII] emission.
This displacement therefore appears to be a real spatial offset of the quasar nucleus with respect to the narrow-line region, presumed to be located at the host galaxy center, further supporting the interpretation that a post-merger gravitational recoil of the SMBH has occurred in E1821+643. 

\end{abstract}




\section{Introduction}

There is compelling evidence that a supermassive blackhole (SMBH) resides at the center of every large galaxy in the universe. 
In active phases, when accreting gas at a sufficient rate, SMBHs can be identified as active galactic nuclei (AGN). 

Galaxy mergers are thought to play a major role in galaxy and SMBH evolution. Following a major merger (two relatively similar sized large galaxies merging), the two SMBHs are expected to form a binary at the center of the merged galaxy \citep{Begelman:1980aa}.
The resulting SMBH binary tightens through dynamical processes and then eventually coalesces through anisotropic emission of gravitational waves (e.g., \citet{Merritt:2005aa}). In general, the gravitational radiation is emitted anisotropically,  causing the merged SMBH to receive a recoil kick  \citep{Favata_2004}. Detailed simulations following the coalescence of spinning black holes {\citep{baker06, Campanelli:2007aa}} have shown that kick velocities up to several $\times 10^3$ kms$^{-1}$ are possible for certain binary black hole (BBH) spin configurations {\citep{Campanelli:2007aa, Lousto:2012aa}}. 

The post recoil oscillations are affected by several different initial conditions including the kick velocity as well as the gas content of the galaxy \citep{Blecha:2016aa}. During the initial harmonic oscillations that are quickly damped due to dynamical friction, the recoiling SMBH can have oscillation amplitudes greater than the radius of the core of the galaxy \citep{Gualandris:2008aa}. These are followed by shorter amplitude, longer lived oscillations which eventually turn to Brownian motion as the SMBH settles down in the center of the merged galaxy.

 The kicked SMBH retains some of the surrounding gas depending on the kick velocity. The radius ($R_{out}$) within which the material remains bound to the recoiling SMBH is given by,
 \begin{equation} 
 R_{out}  \sim \left(\frac{GM}{v^2_{rec}}\right)   \sim  0.43 \left(\frac{M}{10^8 \mbox{M}_\odot}\right)\left(\frac{v_{rec}}{10^3 \mbox{km\,s}^{-1}}\right)^{-2} \,\mbox{pc}
    \label{eq:rout}
 \end{equation}
 
 \noindent where $M$ is the mass and $v_{rec}$ the kick (recoil) velocity of the SMBH \citep{loeb07}. This is the radius at which the Keplerian velocity is equal to the recoil velocity. Depending on the recoil velocity, typically the broad line region (BLR) remains bound to the SMBH while the narrow line region (NLR) is left behind.

 As long as the SMBH continues to accrete gas, it can be identified as an AGN. Recoiling SMBH can then be identified through spatial offsets between the position of the AGN and the center of the galaxy using imaging (e.g, offset optical, IR, Radio, X-ray point sources; \cite[e.g.,][]{Batcheldor:2010aa, Lena:2014aa}). 
A second method of detecting the recoiling SMBH is by analyzing the Doppler shifts of emission lines from the retained gas, showing a relative velocity between the SMBH and the host galaxy. For example, if part of the BLR is retained, the broad Balmer lines will be shifted with respect to the narrow lines \citep[e.g.,][]{Bonning:2007aa}.

Offset AGN are signposts of binary SMBH coalescence, which in turn are strong sources of gravitational waves. Even though the frequencies of these waves are too low to be detected by the Laser Interferometer Gravitational-Wave Observatory (LIGO), identifying these events can help build the base sample for Laser Interferometer Space Antenna (LISA) and Pulsar Timing Arrays. We can also use instances of recoiling SMBH in the nearby universe to further understand the rate of galaxy mergers which allows us to study galaxy evolution.

The quasar E1821$+$643 has been identified as an SMBH recoil candidate by \cite{Robinson:2010aa} (hereafter R10) using spectropolarimetry. E1821$+$643 is a radio-quiet quasi stellar object (RQQ) at a redshift of $z\approx0.3$, and with an absolute
magnitude of M$_V$ = -27.1 \citep{pravdo,hutchings91}, it is one of the most luminous quasars in the
local universe. Although classified as an RQQ, it exhibits many similarities to radio-loud quasars (RLQs) and is associated with an FR-1 radio source \citep{blundellrawl01}.  \citet{hutchings91} have found that the host galaxy is red, featureless and
very large, with an estimated diameter of 75 kpc ($\sim$ 24\arcsec\,), which suggests an elliptical
galaxy -- however most RQQs are located in spiral galaxies. It has a huge extended emission line
region \citep{fried98}, and is located at the center of a cooling flow \citep{russell10}, in a
cluster of galaxies with an Abell richness class $\geq$ 2 \citep{lacy92} -- which again, is usually
typical of RLQs. \citet {russell10} trace the Intra-Cluster Medium (ICM) gas down to $\sim$ 15\,kpc
from the nucleus -- about the extent of the emission line nebula, and calculate that the QSO is
currently accreting at 50\% of the Eddington rate.

\begin{figure} 
\centering 
\includegraphics[width=\columnwidth]{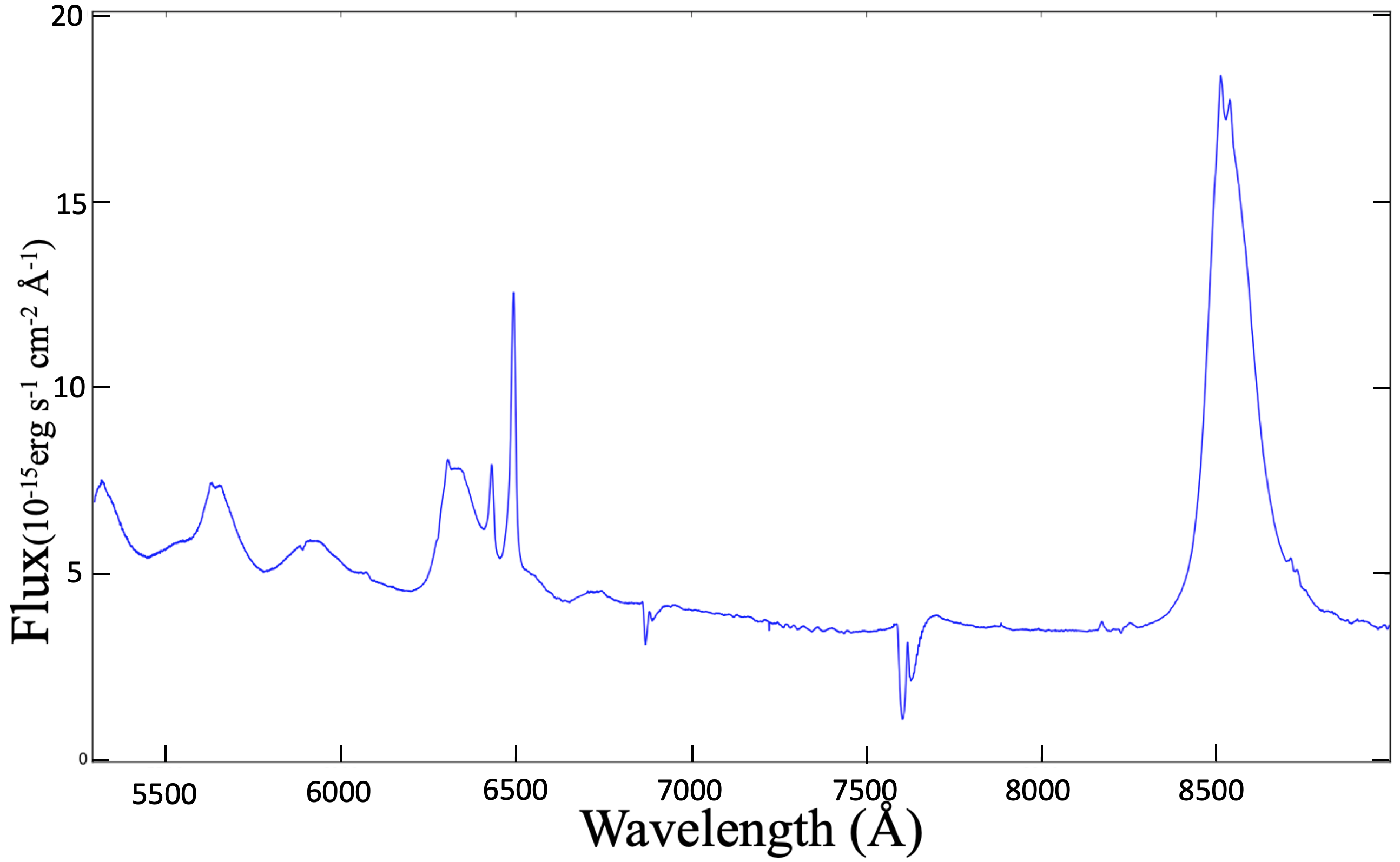} 
\caption{The spectrum of the nucleus
of E1821$+$643 extracted from the combined 2D spectrum obtained with the Gemini Multi-Object Spectrograph.}
\label{spec_all} 
\end{figure}

In the radio, there is extended low-surface brightness emission well beyond the host galaxy
\citep{blundellrawl01}. On arcsecond scales, there is a $\sim$90$^{\circ}$ bend in the south-west jet, which
may be due to precession in a binary, as suggested by \citet{blundell96,blundellrawl01}, but could
also be explained by a spin-flip during a binary supermassive black hole (SMBH) 
merger \citep{merritt02}.

R10 conducted spectropolarimetric observations of E1821+643 as a part of a project to characterize the polarization properties of the broad lines in AGN. They found that the broad Balmer lines are redshifted by $\approx$ 1000 km s$^{-1}$ with respect to the narrow lines and have highly red asymmetric profiles. In polarized light, however, the broad lines are found to be comparably blue shifted with corresponding highly blue asymmetric profiles. R10 explained these characteristics in terms of a scattering model where the BLR moves away from the observer and towards a scattering region in the host galaxy with a speed $\sim 2000$\,kms\,$^{-1}$. 
 Thus, in direct light, the BLR appears to be red shifted with respected to the NLR. However, light from the BLR that is emitted toward the scattering region is scattered back towards the observer, becomes polarized and blue-shifted (as the source is moving towards the scattering screen).

At such a velocity, the accretion disk and the bulk of the BLR remain bound to the moving SMBH. Thus, the spectropolarimetry observations can be interpreted as evidence the SMBH is moving through the galaxy due to a recent gravitational recoil following the coalescence of a progenitor SMBH binary. On the assumption that the $\sim$90$^{\circ}$ bend in the south-west jet is due to a spin-flip,  R10 estimated that the time that has elapsed since the coalescence event is $\sim 1000$\,yr and hence that the recoiling SMBH may have moved a distance $\sim 200$\,pc.

The spectropolarimetry results provide unambiguous evidence of a large relative velocity between the BLR and the host galaxy (as represented by the narrow-line system). However, R10 note that, in addition to the recoil hypothesis, such a velocity difference could also be produced if the BLR forms part of a one-sided wind, or is associated with an active secondary in a binary system.

In an effort to detect the expected spatial displacement and differentiate between the various interpretations, we have obtained high signal-to-noise spectra suitable for spectroastrometric analysis from the Gemini North telescope in Hawaii as well as the Hubble Space Telescope (HST) narrow and broad band imaging observations.

We describe the data reduction, analysis and results in Section \ref{datareduction}, spectroastrometry in Section \ref{spectroastrometry} and HST imaging in Section \ref{hstanalysis}. Simulated spectroastrometry observations were performed to determine if the observed spectroastrometric displacement can be explained by spatially extended [O{\sc iii}] 4959, 5007\AA\ emission (Section~\ref{spectromodelling}). We discuss our results in Section \ref{e1821discussion} and finally, present our conclusions in Section \ref{e1821conclusions}. 

\begin{figure}
\centering 
\includegraphics[width=\columnwidth]{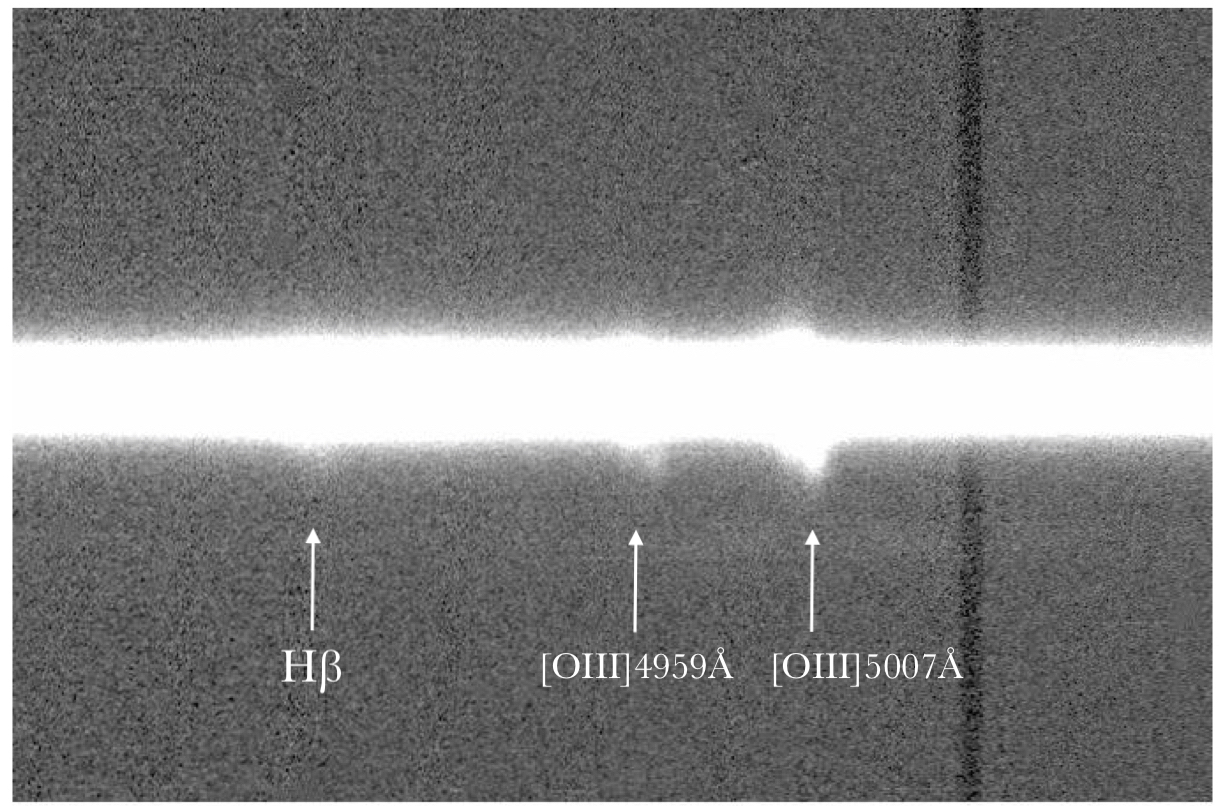} 
\caption{A section of the long-slit spectrum obtained at PA = 0$^{\circ}$, showing the region of H$\beta$ and 
[O{\sc iii}] 4959, 5007\AA\, (in order from left to right). The extended [O{\sc iii}] can clearly be seen towards the north 
(bottom of the image). The detector chip-gap can also clearly be seen to the right of the [O{\sc iii}] lines.}
\label{2d_spectra} 
\end{figure}

At the redshift, $z=0.297$, of E1821+643, the angular scale is 4.568 kpc/arcsec, assuming $H_0 = 67.4$\,km\,s$^{-1}$\,Mpc$^{-1}$, and $\Omega_m = 0.315$\citep{Planck2020}.

\section{Observations and Results}

\label{datareduction}
\subsection{Spectroastrometry}
\label{spectroastrometry}
\subsubsection{Observations}

\begin{figure}
\centering 
\includegraphics[width=\columnwidth]{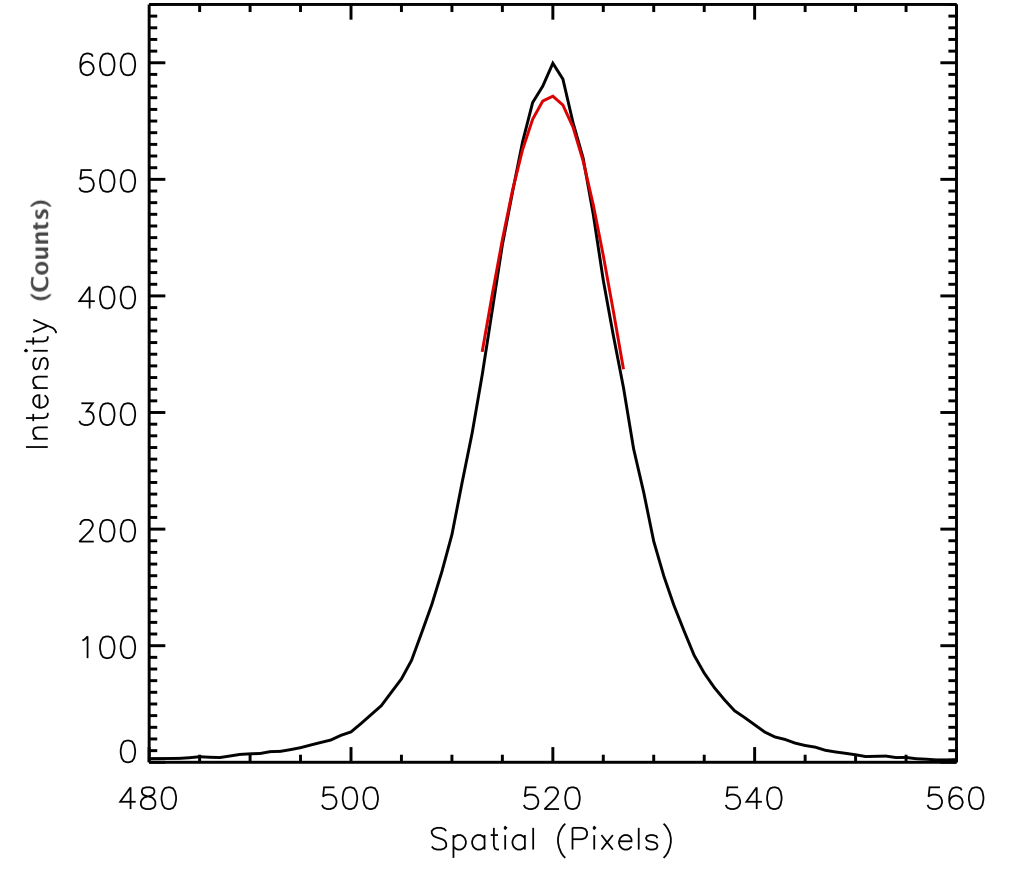} 
\caption{ The spatial profile of the [O{\sc iii}] 5007\AA\ emission line (black), extracted from the PA=0$^{\circ}$ spectrum at the line center wavelength ($6493.2$\AA). The red line  represents a gaussian fit to the spatial profile. The fit is restricted to the peak of the profile, between the $25^{th}$ and $75^{th}$ percentiles, to prevent any extended emission from biasing the fit.}
\label{spatial_profile} 
\end{figure}

We acquired deep optical spectroscopy of E1821$+$643 in an attempt to detect a spatial offset between the broad and narrow emission lines by means of spectroastrometric analysis, i.e. studying the spatial displacements between the different emission line velocity
systems \citep[see ][for a description of the spectroastrometry technique]{bailey98}. The observations were obtained at the Gemini North
telescope in Hawaii, using the Gemini Multi-Object Spectrograph (GMOS) in long-slit mode, on the nights of 20100424 (science) and 20100323 (flux calibrations). The R400 grating was used in conjunction with a 1\arcsec\, slit centered on the quasar nucleus. Two exposures 
of 600\,s each were obtained with the slit at position angles of 0$^{\circ}$,
90$^{\circ}$ and 270$^{\circ}$. See figure \ref{2d_spectra} for a section of the spectra at 0$^{\circ}$. Due to a ``seeing bubble'', 3$\times$600s spectra were taken with the
slit at a position angle of 180$^{\circ}$, although only one of these was deemed useful for the
spectroastrometry analysis. CuAr arcs were taken at each position angle, immediately prior to the
object spectra.

\subsubsection{Data Reduction \& Analysis}
\label{reduction_analysis}

The data were reduced using the {\em Gemini: GMOS} package within {\em IRAF}. This was used to
subtract the bias, divide by a normalized flatfield, combine the data from the three CCDs, remove
the sky emission lines and, wavelength and flux calibrate the two-dimensional spectra. The spectra
were then trimmed to remove second order contamination at the red wavelength end and
cosmic rays were removed in the standard way.

Combined 2-D spectra were formed for the North-South and East-West directions. The two exposures at each position angle were first co-added, with the exception of PA = 180$^{\circ}$, for which only one usable spectrum was obtained. The usable spectrum taken at PA=180\degr\, was aligned and combined with the averaged PA = 0\degr\, spectrum, to produce a single North-South spectrum with a total exposure time of 1800\,s. Similarly, the co-added PA = 270\degr\, and 90\degr\, spectra were 
aligned and combined to produce a single East-West spectrum with a total exposure time of 2400\,s .
One-dimensional spectra were then extracted from 1\arcsec\, apertures in 1\arcsec\, spatial increments along each direction.  

Finally, all of the 2-D spectra were combined to form a single spectrum with a total exposure time of 4200\,s, from which a 1-D spectrum of the nucleus was then
extracted using a 1\arcsec\, aperture. The resultant spectrum can be seen in figure~\ref{spec_all}.

\begin{figure*} 
\centering 
\includegraphics[width=\textwidth]{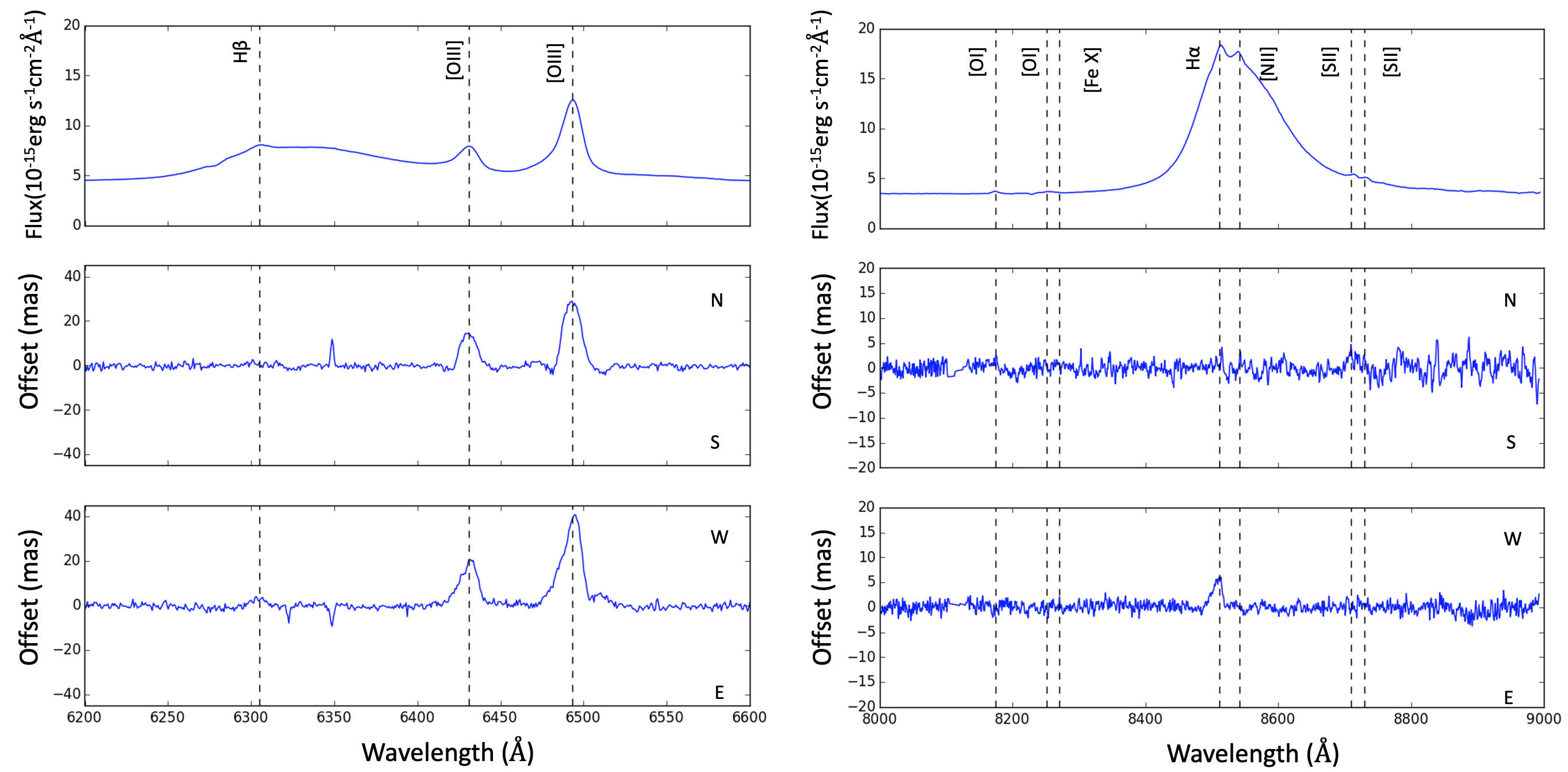} 
\caption{The displacement spectra derived from the Gemini long-slit observations. The dotted lines indicate the emission lines. The top panel is the total flux. The middle and bottom panels show the displacement spectra in the N-S and E-W direction respectively. } 
\label{fig:cont_fit} 
\end{figure*}

\begin{figure*} 
\centering 
\includegraphics[width=0.8\textwidth]{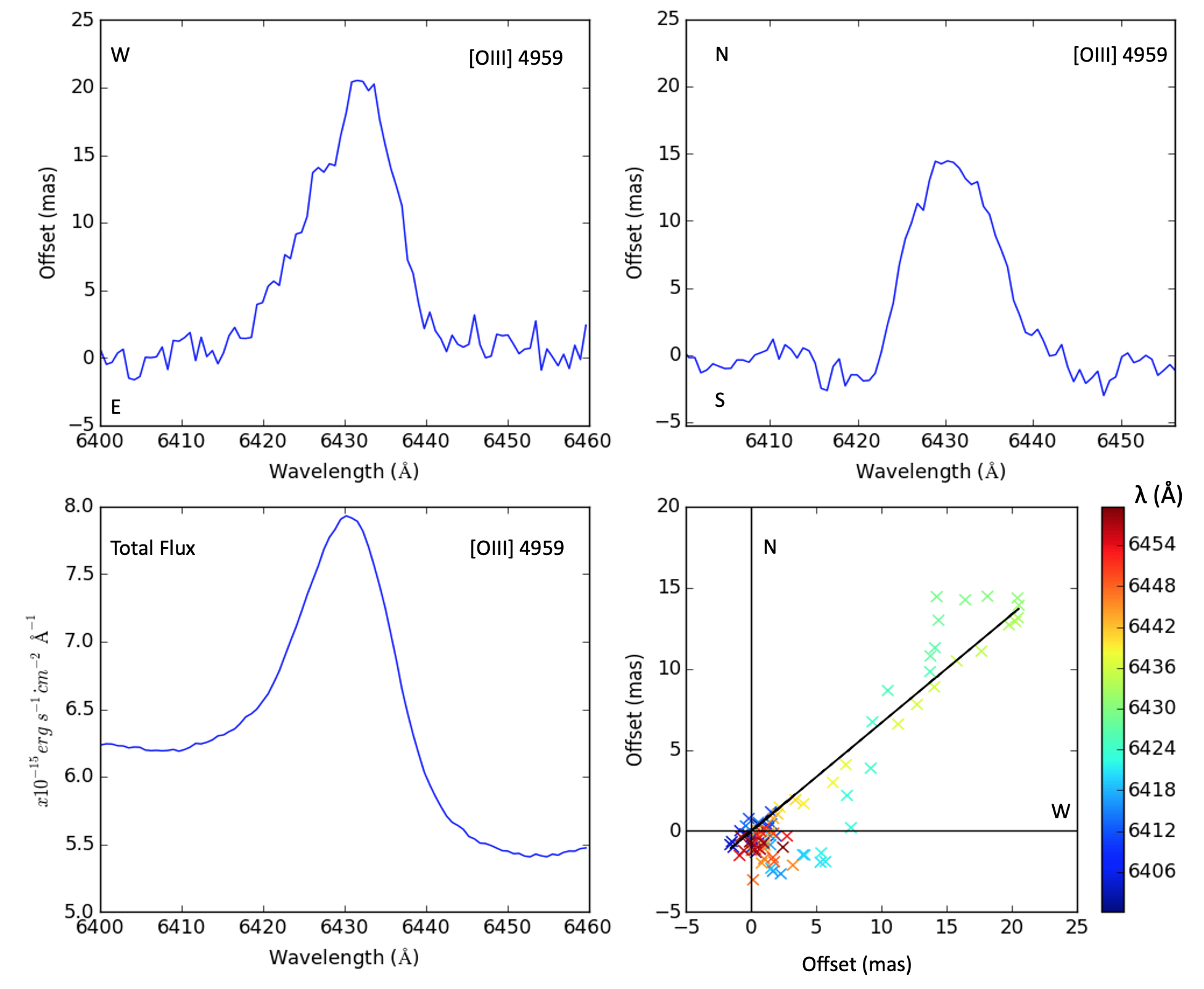} 
\caption{Displacement and flux profiles for [OIII] 4959\AA . The top left and right panels show the offset of the [O{\sc iii}]  line in the E-W and N-S directions. The bottom left panel shows the flux for the [O{\sc iii}] 4959\AA\ line. The bottom right panel shows the N-S and E-W offset components of the [O{\sc iii}] 4959\AA\ line as color coded by wavelength along with the best fit line (in black).} 
\label{fig:dispo34959} 
\end{figure*}

\begin{figure*} 
\centering 
\includegraphics[width=0.8\textwidth]{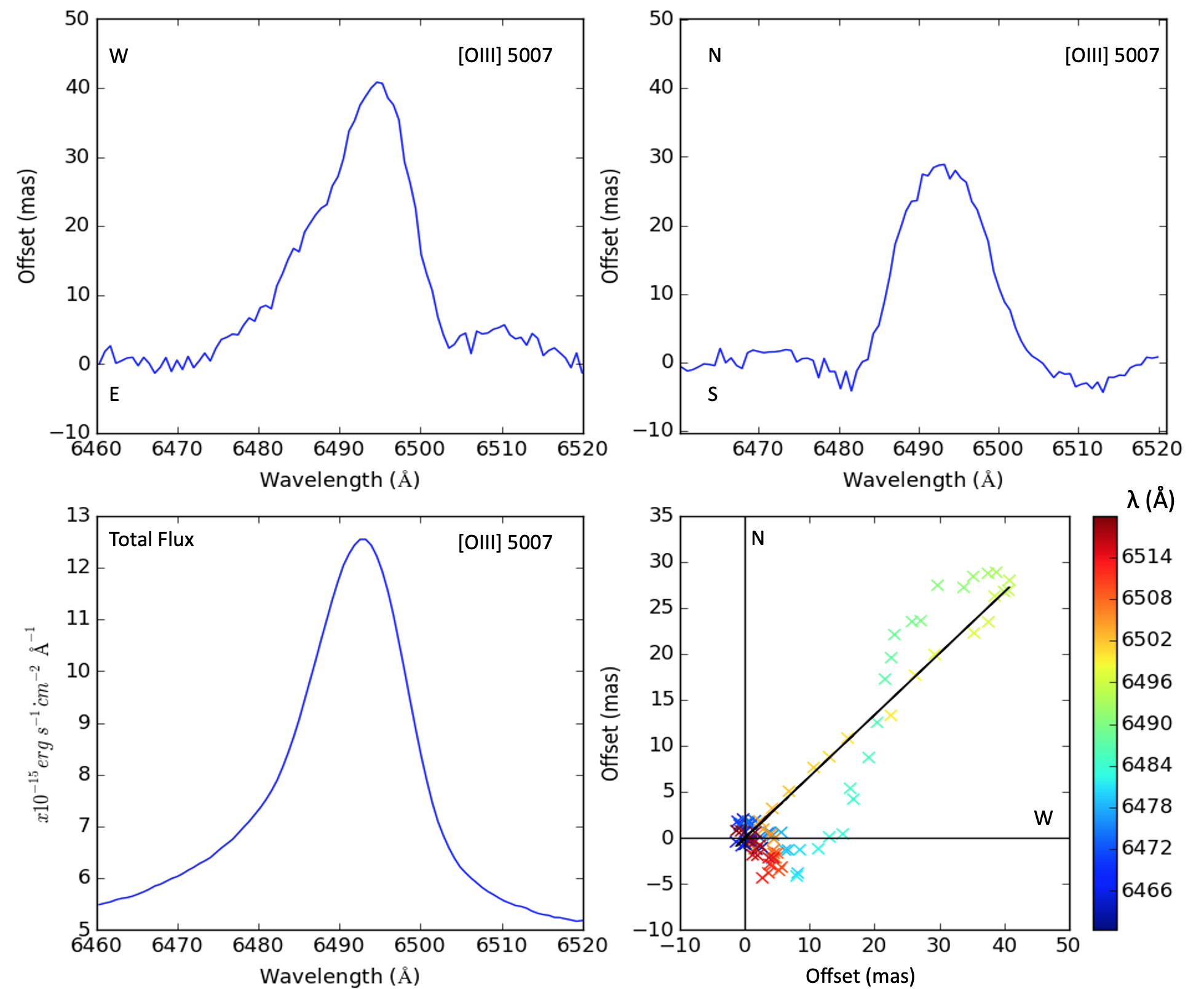} 
\caption{As figure \ref{fig:dispo34959} for [O{\sc iii}] 5007\AA}
\label{fig:dispo35007} 
\end{figure*}

Spectroastrometry is a technique for measuring the relative spatial displacements of spectral features such as emission lines. This is done by measuring the position of the source in the long slit 2-D spectrum as a function of wavelength. If the spatial profile is sufficiently well sampled, the position of its centroid can be measured to a precision of the order of milli-arcseconds. Any wavelength dependent asymmetries in the spatial profile will appear as deviations in the resulting `displacement spectrum'.  

 Spectroastrometry was carried out on the co-added 2D spectra obtained at each PA following the general method described by \cite{bailey98}. Gaussians were fitted to the spatial profiles to determine the position of the peak of the flux distribution for each pixel along the dispersion direction. However, as our HST images (Section~\ref{hsto3}) show that the nucleus is surrounded by extended [O{\sc iii}] $\lambda \lambda$4959,5007 emission, the spatial profiles were fitted between the $25^{th}$ and 75$^{th}$ percentiles (see Fig.~\ref{spatial_profile}), to minimize any bias in the peak position due to asymmetries in the spatial distribution of the extended line emission. The ''spectrum'' of peak positions derived from the fits was then fit using a low-order polynomial in wavelength to remove overall curvature in the `continuum' spectra. The spatial offsets from the fitted polynomial were then calculated at each wavelength, yielding a spectrum of the displacement relative to continuum for each PA. Finally, the PA =180$^{\circ}$ data were  subtracted from the 0$^{\circ}$ data, and the 270$^{\circ}$ data were subtracted from the 90$^{\circ}$ data to remove any instrumental effects. The final result is a spectrum of the spatial displacements in, respectively, the North-South and East-West directions. 

\subsubsection{Spectroastrometry results}
\label{sec:specast}

In this case, we are looking for a displacement between the broad and narrow emission lines in the spectrum of E1821+643.  
Such spatial displacements are expected in the recoil hypothesis if the BLR remains bound to the SMBH but the gas emitting the narrow lines is left behind.
Segments of the displacement spectra covering H$\beta$ -- [O{\sc iii}] and H$\alpha$ -- [N{\sc ii}] are shown in figure~\ref{fig:cont_fit} for both the N-S and E-W directions. The redshifted wavelengths of the emission lines that fall within these segments are indicated. The displacement is measured relative to the continuum. As this is dominated by the quasar nucleus, it has zero displacement by definition. Similarly, the broad components of the H$\alpha$ and H$\beta$ lines show no displacement relative to the continuum. This is not surprising since the BLR is expected to have an angular size $\lesssim 1$\,mas, which is not detectable in our spectroastrometry data.  The N-S and E-W displacement spectra show clear offsets in the [O{\sc iii}] $\lambda \lambda$4959,5007 lines in the North and West directions. Smaller, but still significant displacements are also associated with the narrow components of H$\alpha$ and H$\beta$. The other lines do not show any significant offset.

 The amplitudes of the displacements in the [O{\sc iii}] lines  were determined by fitting gaussians to the peaks of the displacement profiles. This yields amplitudes of $\approx 40$ and 30\,mas to the West and North of the nucleus, respectively, in [O{\sc iii}] $\lambda 5007$ (Table~\ref{o3disp_table1}). The maximum displacements in [O{\sc iii}] $\lambda 4959$ are about half those in [O{\sc iii}] $\lambda 5007$. The narrow components of H$\alpha$ and H$\beta$ only show significant displacements to the West of the nucleus, with amplitudes $\sim 5$ and 3\,mas, respectively. No significant displacements are seen in the other narrow lines, such as [O{\sc i}] $\lambda\lambda 6300, 6363$, [N{\sc ii}] $\lambda\lambda 6548, 6583$, and [S{\sc ii}] $\lambda\lambda 6717, 6731$, which are, however, comparatively weak.

\begin{table}
\label{o3disp_table1}
\centering
\caption{Measured displacements for the [O{\sc iii}]  lines}
\begin{tabular}{lccc}
\hline
\hline
Line & PA &$\mu_{NS}$ &$\mu_{EW}$ \\
     & $^{\circ}$ & mas & mas \\
\hline

[OIII] 4959 \AA\, & -52.9 &14.9$\pm$1.0 &19.7$\pm$1.9   \\

[OIII] 5007 \AA\, & -53.0 &29.5$\pm$2.4 &39.2$\pm$1.2  \\

\hline
\end{tabular}

The values {$\mu_{NS}$  and $\mu_{EW}$ were derived by fitting Gaussians to the displacements in the North-South and East-West direction. The uncertainties are the 1-$\sigma$ errors on the best fit amplitude. The PAs are calculated from the displacement components.}
\end{table}

The displacements across the [O{\sc iii}] $\lambda \lambda$4959,5007 lines are shown in more detail in figures~\ref{fig:dispo34959} and \ref{fig:dispo35007}, respectively. 
In addition to having different amplitudes in the E-W and N-S directions, the displacement profiles show distinct structures. The flux profiles are asymmetric, showing an extended blue wing in the E-W direction. The corresponding displacement profile is similarly asymmetric, showing a prominent extended wing on the blue side of the profile, which indicates a displacement to the West. In the N-S direction, the cores of both [O{\sc iii}] lines show large displacements to the North, as already noted, but the wings exhibit small displacements ($\sim 5$\,mas in [O{\sc iii}] $\lambda$5007) to the South on both sides of the profile.
The [O{\sc iii}] 4959\AA\, line shows similar structure, albeit with smaller displacements.

The E-W and N-S components of the measured displacement are plotted in the lower right panels of figures~\ref{fig:dispo34959} and \ref{fig:dispo35007}. The corresponding radial displacement in [O{\sc iii}] $\lambda$5007 has a magnitude $\approx 50$\,mas with a position angle PA$\approx 305^{\circ}$. However, the measured displacements are affected by wavelength-dependent ''dilution'' due to the underlying quasar light, which has a greater effect on the weaker lines. For example, both of the [O{\sc iii}] lines are superposed on the very extended red wing of the broad H$\beta$ line (Figure~\ref{spec_all}), but [O{\sc iii}] $\lambda$4959 is not only weaker than [O{\sc iii}] $\lambda$5007 but the local quasar flux (i.e., the continuum and broad H$\beta$ emission) is also higher and therefore its displacement signal suffers relatively higher dilution. A simple correction for this dilution effect is presented in Section~\ref{contcorr}.

The curved pattern in the displacement locus seen in the bottom right panels in figures \ref{fig:dispo34959} and \ref{fig:dispo35007} is a result of the asymmetry in the E-W flux profile, which is not seen in the N-S displacement profile. This asymmetry appears to be associated with the extended blue wing in the flux profiles, suggesting that there is an outflow present, which is predominantly to the west of the nucleus.

\subsection{HST imaging}
\label{hstanalysis}

\subsubsection{HST observations}
 Understanding the morphology of the narrow line region on sub-arcsecond scales will help to determine whether the spectroastrometric displacements presented in Section~\ref{sec:specast} are due to a spatial offset between the quasar nucleus and the NLR or an asymmetric distribution of the [O{\sc iii}]  emission caused by partial extinction of a bi-polar outflow. In order to map the [O{\sc iii}] emission around the quasar and the host galaxy we  obtained HST ACS (Advanced Camera for Surveys) Wide Field Camera images of E1821+643 (proposal ID 13385, PI A. Robinson). Narrow band images centered on the [O{\sc iii}]$\lambda \lambda$4959,5007 and H$\beta$ lines were obtained using the FR656N ramp filter set to central wavelengths of 6460\AA and 6305\AA, respectively. Medium band continuum images were also obtained using the F647M ramp filter centered at 7300\AA. 

For the [O{\sc iii}] and continuum observations, we used a combination of short exposures to obtain high signal-to-noise but unsaturated images of the quasar nucleus and longer exposures to map the host galaxy and extended [O{\sc iii}] emission on scales $\sim$ a few arcseconds. A two-point dither pattern was used for each set of exposures.

\begin{figure}  
	\centering
	\includegraphics[width=\columnwidth]{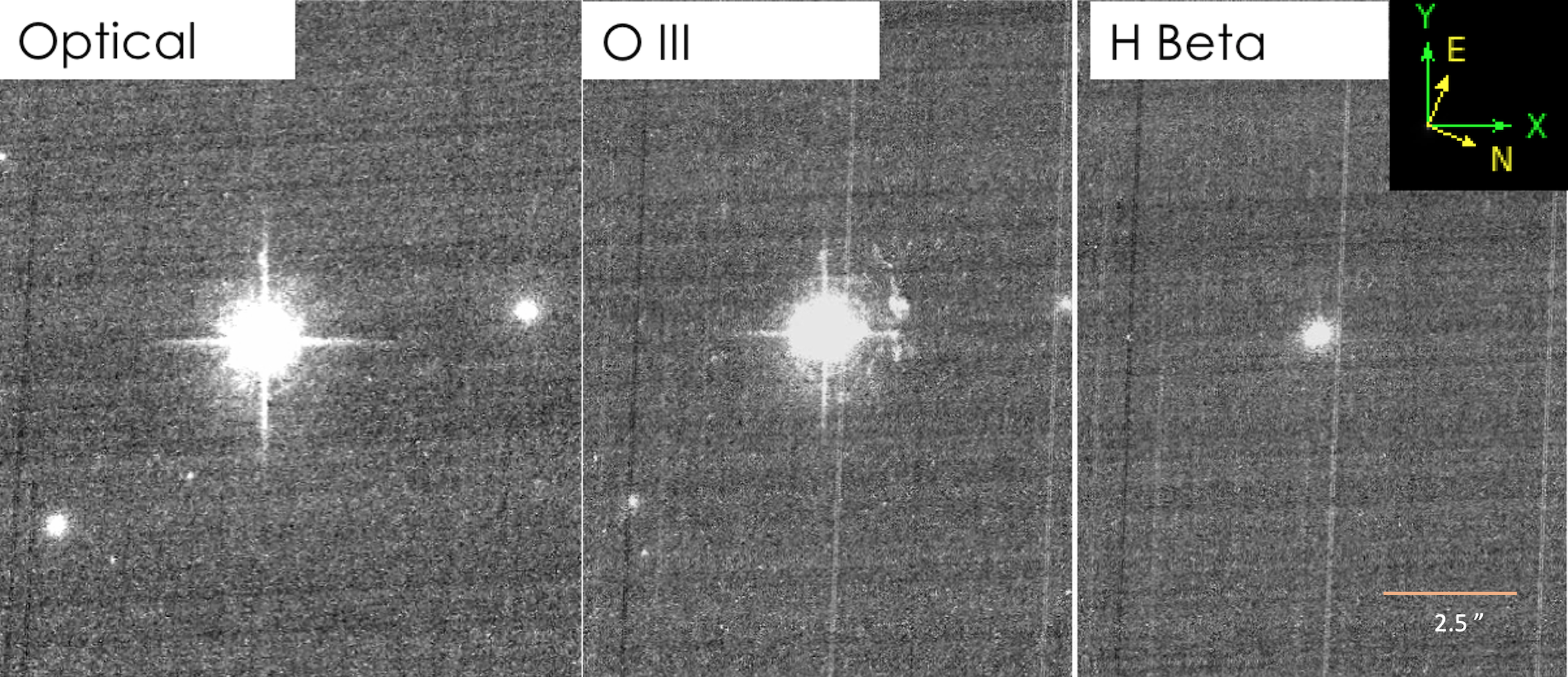}
	\caption{HST ACS/WFC images. Left to right, FR647M, FR656N, FR656N.}
	\label{fig:hstimg}
\end{figure}

The dithered images were processed and combined through the ACS data calibration pipeline. After processing, the saturated pixels around the quasar nucleus in the long exposure images were replaced with the corresponding unsaturated pixels extracted from the the short exposure images and scaled to the long exposure image. The resulting images are shown in figure \ref{fig:hstimg}.

Visual inspection shows that the quasar nucleus is the dominant source in both the continuum and [O{\sc iii}] images. In the continuum image, only the quasar point spread function (PSF) is visible, showing that the host galaxy contribution is negligible. The quasar PSF is also the dominant feature in the narrow band image. However, clumpy emission extending in an arc from North to North-East is visible at a distance $\sim 3$ arcseconds (13.7 kpc) from the nucleus. This is not present in the continuum image and can therefore be attributed to the [O{\sc iii}] lines.
However, in order to map the [O{\sc iii}] on smaller scales, it is necessary to subtract the quasar PSF. 

\subsubsection{HST image analysis}
\label{Sec:hstdecomp}
PSF subtraction was performed for the FR656N image both manually and using the \texttt{Galfit} image decomposition software, which models the 2D light distribution of galaxy images \citep{Peng:2002aa}. An ACS image of a white dwarf standard star, GRW+70D5824 (spectral type DA3), observed with the same filter FR656N was obtained from the HST Legacy Archive (proposal ID: 9563) and used as a model for the PSF.  

To facilitate the PSF modeling we re-sampled the image, to expand the pixel dimensions of the image by a factor of 3. The up-sampled image has a plate scale of 0.017\arcsec\,/pixel. 

\begin{figure}
	\centering
	\includegraphics[width=\columnwidth]{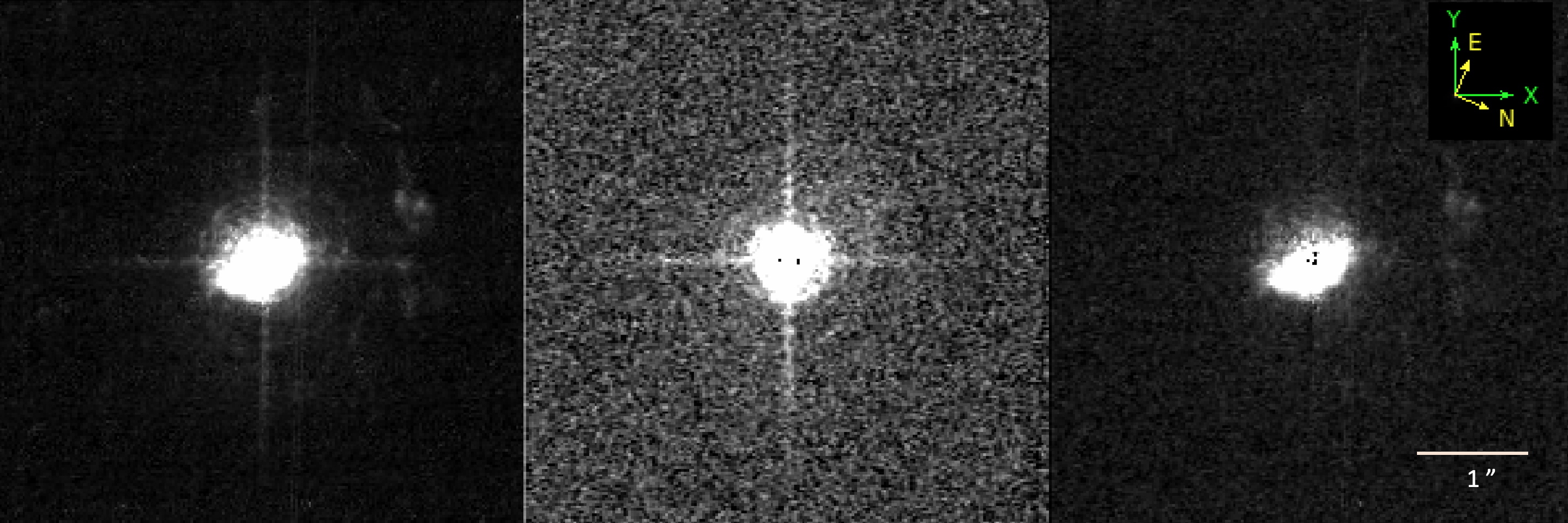}

	\caption{Manual PSF subtraction. From left, the first panel shows the FR656N ramp filter [OIII] image, the middle panel shows the standard star used as the PSF template and the last panel shows the asymmetric distribution of the [OIII] emission after subtraction of the point source.
	The direction is the same as in figure \ref{fig:hstimg}. The individual images have been scaled in brightness in order to highlight the circum-nuclear morphology.}
	\label{fig:psfmanual}
\end{figure}

For the manual subtraction, the flux of the standard star was scaled to match that of the nuclear point source in the FR656N [O{\sc iii}] image. However, to avoid  over-subtraction of the [O{\sc iii}] emission it is necessary to account for its contribution to the flux of the unresolved quasar nucleus. We estimated the [O{\sc iii}] contribution by fitting the [O{\sc iii}]$\lambda \lambda$4959,5007 lines in the Gemini spectrum of the nucleus of E1821+643 (Fig.~\ref{spec_all}). The fit was used to generate a spectrum containing only the [O{\sc iii}] lines and both this and the nuclear spectrum were separately convolved with the FR656N filter transmission curve generated using the formulae provided by \cite{Bohlin:2000} in order to estimate the fraction of the the nuclear flux within the FR656N bandpass that is due to [O{\sc iii}] emission. We find that about $\sim$20\% of the nuclear flux in the 1\arcsec\ aperture from which the spectrum was extracted is contributed by the [O{\sc iii}] lines. In order to perform the PSF subtraction, the integrated flux of the standard star (i.e. the PSF template) within an aperture of 1\arcsec\ was matched to that within the same sized aperture centered on the nucleus in the FR656N [O{\sc iii}] image and then scaled by 80\% to account for the [O{\sc iii}] contribution. The star was then subtracted from the [O{\sc iii}] image to reveal the distribution of the resolved [O{\sc iii}] emission as seen in figure \ref{fig:psfmanual}.

\begin{figure}
	\centering
	\includegraphics[width=\columnwidth]{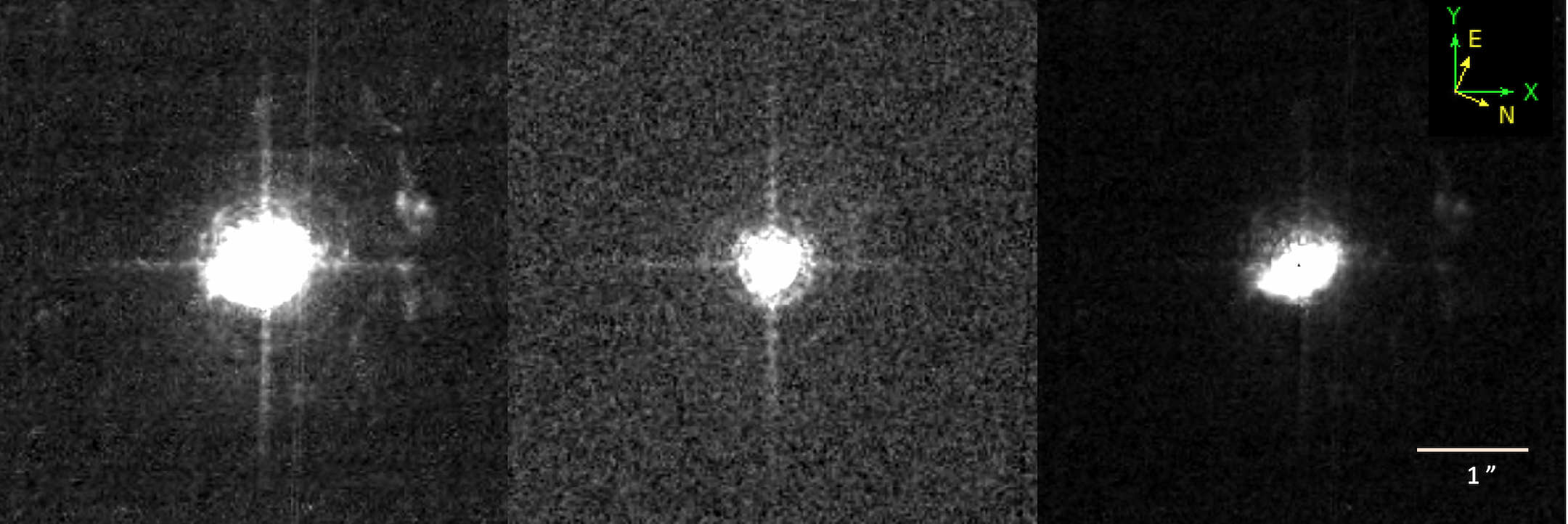}

	\caption{PSF subtraction using \texttt{Galfit}. From left, the first panel shows the FR656N ramp filter [OIII], the middle panel shows the \texttt{Galfit} model and the last panel shows the \texttt{Galfit} residual showing the distribution of the [OIII] emission after subtraction of the point source.}
	\label{fig:psfgalfit}
\end{figure}

\texttt{Galfit} was also used to fit the quasar nucleus in the FR656N [O{\sc iii}] image, using the scaled standard star as the PSF template (middle image in figure \ref{fig:psfgalfit}). A S\'ersic component was also included in the fit to model the galaxy light. which was then subtracted from the original to reveal the [O{\sc iii}]  distribution. 
However, no significant differences were found in the residuals between models which included or excluded the S\'ersic component  component, suggesting that the galaxy component is negligible in comparison with the quasar nucleus.

\begin{figure}
	\centering
	\includegraphics[width=\columnwidth]{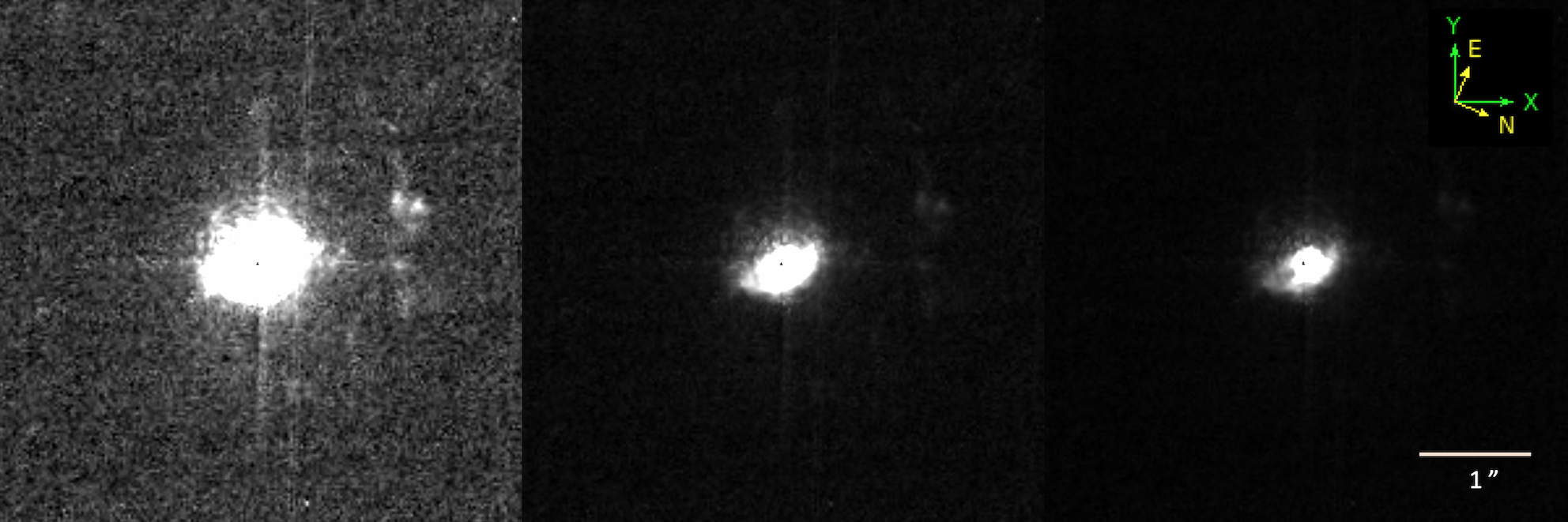}

	\caption{FR656N ramp filter [O{\sc iii}] image after manual PSF subtraction (Fig.~\ref{fig:psfmanual}). The image is shown at three different contrast levels to reveal the morphology of the circum-nuclear and extended [O{\sc iii}] emission around the nucleus of E1821+643.}
	\label{fig:e1821slice}
\end{figure}

\begin{figure}
	\centering
	\includegraphics[width=\columnwidth]{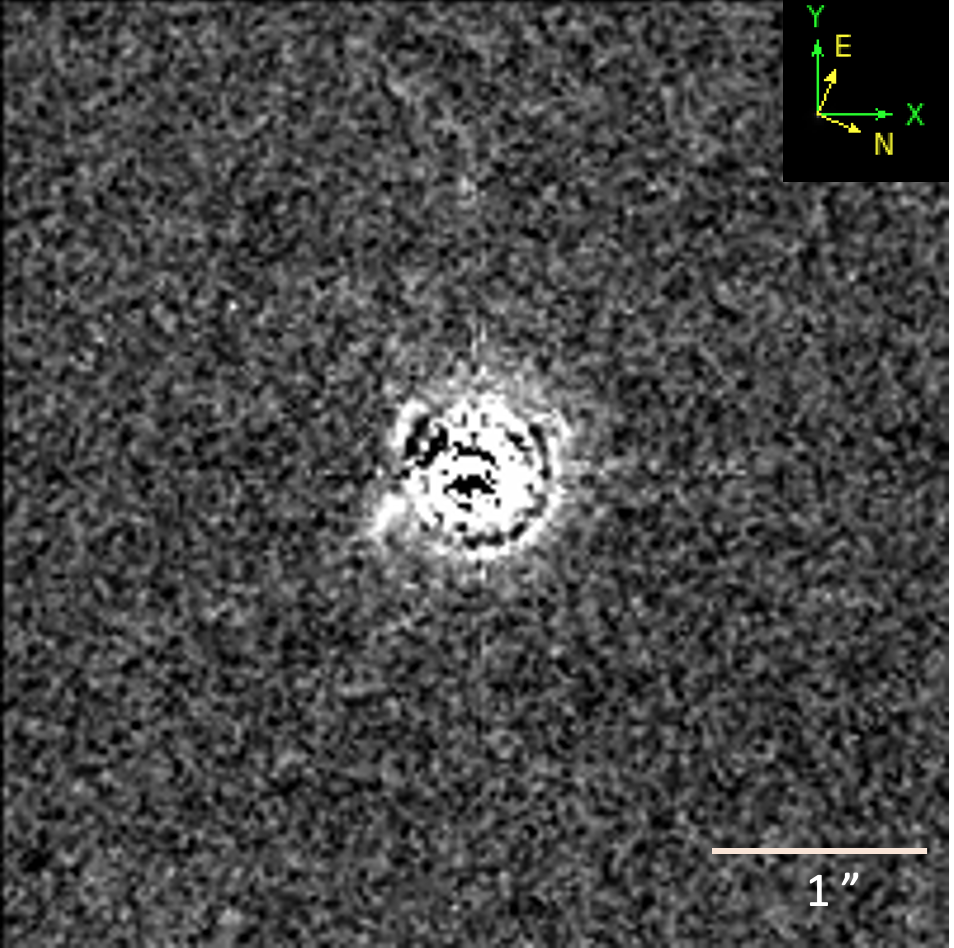}

	\caption{Residual of the ACS/WFC F647M continuum image of E1821+643 after point source subtraction.}
	\label{fig:contres}
\end{figure}

Although visual inspection of the F647M (continuum) image, as well as  modeling of the FR656N  [O{\sc iii}] image, suggests that the host galaxy contribution is insignificant, we nevertheless attempted to quantify this by modeling the continuum image with \texttt{Galfit}. An ACS image of a white dwarf standard star, GD71, observed with the same filter, FR647M, was obtained from the HST Legacy Archive (proposal ID: 9020). The unresolved quasar nucleus was fitted using the star as a PSF model. The residual image, after subtraction of the fitted PSF model is shown in fig \ref{fig:contres}. As can be seen, PSF structure is still present in the residual image, showing that the quasar light has not been entirely removed. We performed aperture photometry on the original continuum image and on the residual image, to obtain relative fluxes within the three regions used for the simulated spectroastrometry observation described in Section~\ref{spectromodelling} (the unresolved nucleus within a radius $0.12\arcsec$,  the annulus between 0.12$\arcsec$ and 0.5$\arcsec$ and the surrounding region beyond 0.5$\arcsec$). We found that the residual image only contributes $\sim10\%$ of the total light in the continuum image within all three regions. However, due to the PSF mismatch, the residual contains a significant proportion of light from the quasar, and so the flux measured from residual image only sets an upper limit on the contribution of the galaxy component. This result is consistent with the image decomposition analyses of \cite{Floyd2004} and \cite{Kim:2017}, both of whom estimate that the {\em entire galaxy} contributes $\sim10\%$ of the total flux in the $I$ band.

\subsubsection{Results of HST image analysis}
\label{hsto3}

Both the manual subtraction as well as the \texttt{Galfit} modeling produced similar spatial distributions of the [O{\sc iii}]  emission after PSF subtraction as seen in figures \ref{fig:psfmanual} and \ref{fig:psfgalfit}. The spatial distribution of the [O{\sc iii}]  emission is shown at three different contrast levels in Figure \ref{fig:e1821slice}, with the left panel showing the extended structure while the other two panels reveal the morphology of the circum-nuclear emission.
It is apparent that the emission around the nucleus is partly resolved and forms a wide fan-shaped structure, spanning approximately $135\degr$ of azimuth from North-East to West. This structure is approximately 1\arcsec\ (4.6\,kpc) wide and 1.6\arcsec\ (7.3\,kpc) long. The emission extends radially as far as $\sim 0.5$ arcsecond North of the quasar nucleus, which is slightly off-center to the South-East. There is a second, fainter, arc-like structure of clumpy [O{\sc iii}]  emission that extends from East to North at distances $\sim 3$\arcsec\ (13-14\,kpc) from the nucleus. This feature is probably photoionized by the quasar and may be associated with a tidal tail from on ongoing interaction with a gas rich satellite galaxy (Rosborough et al., 2021, in preparation).

\section{Spectroastrometry modelling}
\label{spectromodelling}

In this section, we present a simulated spectroastrometry observation, to determine if the measured displacements can be explained by the asymmetric distribution of the circum-nuclear extended [O{\sc iii}] emission, and describe a simple model to correct the measured displacements in the [O{\sc iii}] lines for dilution by the quasar continuum and broad H$\beta$ line.

\subsection{Simulated spectroastrometry observation}
\label{specastobsims}

\begin{figure}
	\centering
	\includegraphics[width=\columnwidth]{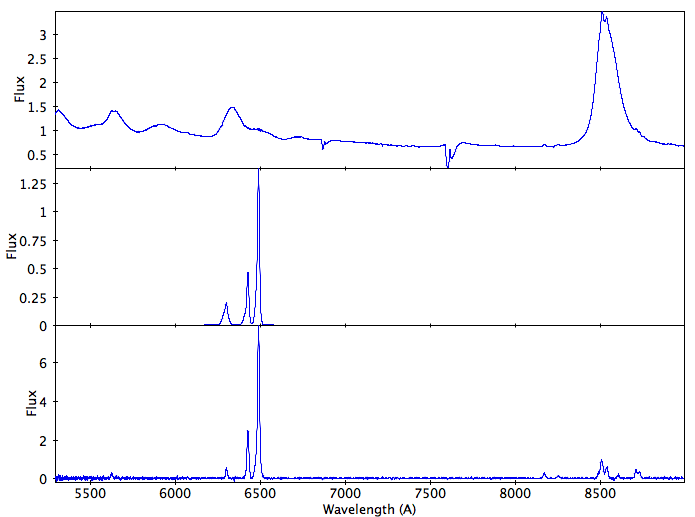}
	\caption{Spectra used in the spectroastrometry modeling. The top panel shows the spectrum extracted from a 1 arcsec aperture centered on the nucleus. The [O{\sc iii}] and narrow H$\beta$ lines have been fitted and subtracted so that this spectrum shows only the quasar continuum and the broad emission lines. The middle panel shows the model fitted to the [O{\sc iii}]  lines. The bottom panel shows the spectrum extracted from a 1$\arcsec$ aperture offset 1\arcsec\ from the nucleus. All plots show the flux density normalized to the flux obtained by convolving the spectrum with the FR656N filter response. }
	\label{fig:offsetspectra_all}
\end{figure}

\begin{figure}
	\centering
	\includegraphics[width=\columnwidth]{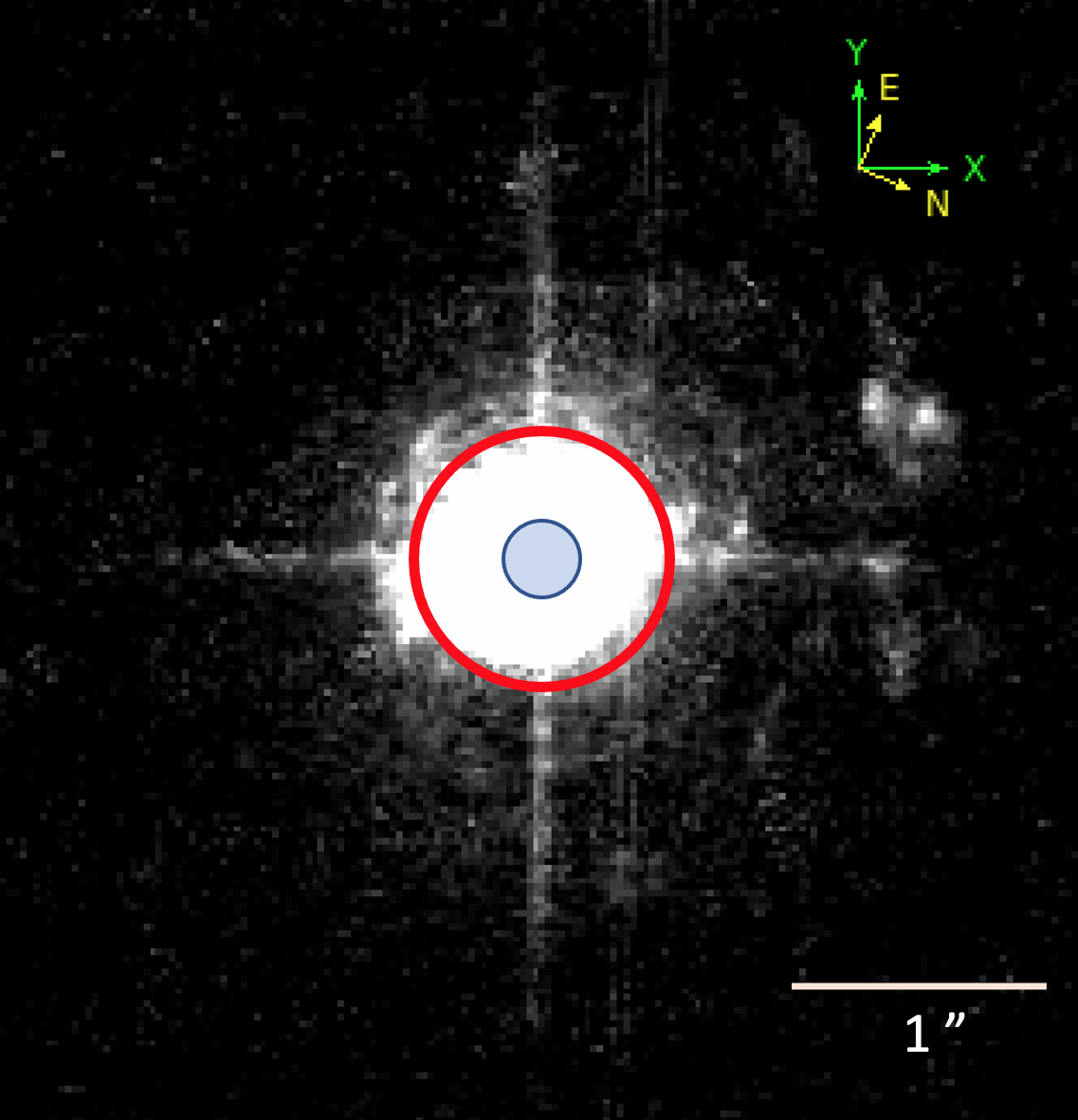}
	\caption{ Annuli used for the spectroastrometric simulations superimposed on the FR656 [OIII] image. The inner blue circle shows the unresolved nucleus enclosing the PSF within the first Airy ring. This region is represented by a spectrum that includes the quasar continuum and broad line emission with a 90\% contribution from the [O{\sc iii}] lines, representing the unresolved [O{\sc iii}] emission (see text and Fig.~\ref{fig:offsetspectra_all}, top and middle panels). The annulus between the blue and red circles encloses the [O{\sc iii}] emission that is resolved in the HST image but not in the Gemini spectra. This region is represented by the fitted narrow lines extracted from the spectrum of the nucleus (Fig.~\ref{fig:offsetspectra_all}, middle panel). Beyond the red circle, the extended [O{\sc iii}] emission is represented by a spectrum extracted from an aperture offset by 1$\arcsec$ from the nucleus (Fig.~\ref{fig:offsetspectra_all}, bottom panel). The blue and red circles have radii of 0.12$\arcsec$ and 0.5$\arcsec$, respectively.}
	\label{fig:spec_rings}
\end{figure}

\begin{figure}
	\centering
    \includegraphics[width=\columnwidth]{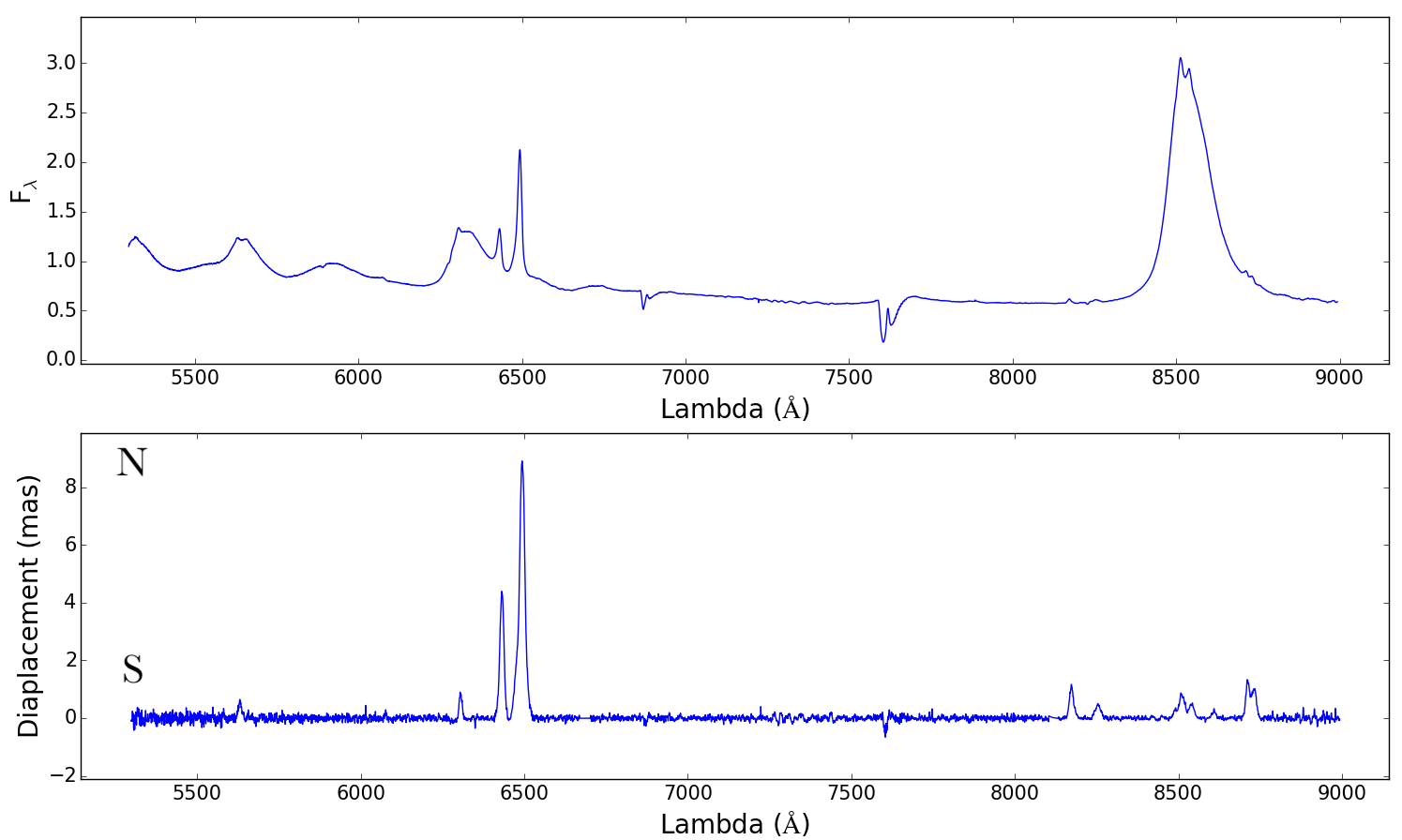}\\
    \includegraphics[width=\columnwidth]{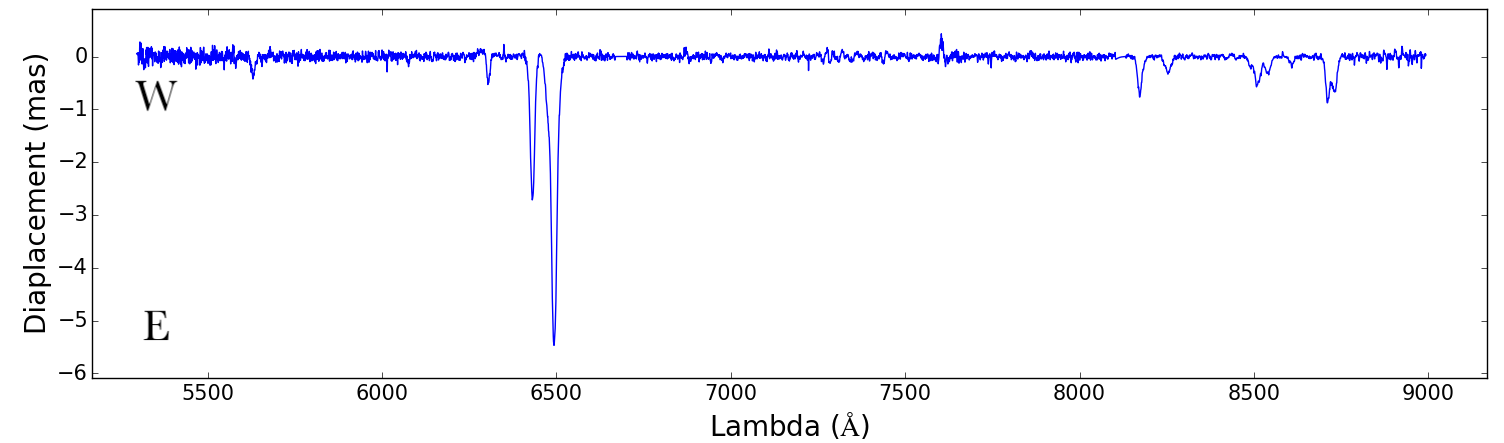}
	\caption{Spectroastrometric modelling results: The top panel shows the reconstructed spectrum of the quasar extracted from the spectroastrometry modeling simulations. The middle and bottom panels show the simulation of spectroastrometric displacement spectrum where you see the expected displacement. Middle panel: the slit is along North-South Direction. Bottom panel: Slit is along West - East direction }
	\label{fig:spec2}
\end{figure}

The spectroastrometry shows a significant displacement between the [O{\sc iii}] lines and the nucleus in the North-West direction (Section~\ref{sec:specast}) . However, as discussed above (Section~\ref{hsto3}) the HST image analysis shows that the [O{\sc iii}] emission is also asymmetrically distributed with respect to the nucleus, in broadly the same direction as the spectroastrometric displacement. Thus, we must consider the possibility that part or all of spectroastrometric displacement is actually caused by the asymmetric spatial distribution of the [O{\sc iii}] emission rather than a spatial offset of the quasar nucleus. To test this, we constructed a simulated spectroastrometry observation by combining the HST [O{\sc iii}] image with spectra of the nucleus and the extended [O{\sc iii}] emission extracted from the Gemini data.

A three dimensional array was created by combining the re-sampled [O{\sc iii}] image with the Gemini spectra, where the third dimension is wavelength.  Each pixel of the image was assigned one of the three spectra described below, depending on its position. Each spectrum was first normalized by the flux obtained by convolving the spectrum with the FR656N filter response and then scaled to the pixel flux from the image. The three spectra represent, respectively, the unresolved nucleus in the HST image, an annulus surrounding the unresolved nucleus of outer radius 0.5$\arcsec$, which corresponds to the 1$\arcsec$ aperture from which the Gemini spectrum of the nucleus was extracted, and the extended [O{\sc iii}] emission at distances $> 0.5$$\arcsec$ from the nucleus. These regions are indicated in figure \ref{fig:spec_rings} by the blue shaded circle and the red circle, respectively. 
The region containing the unresolved nucleus (blue circle in Figure \ref{fig:spec_rings}) is defined by a circle centered on the peak of the PSF, which has a radius of 0.12$\arcsec$ ($\sim$7 re-sampled pixels). This is approximately the radius of the first Airy ring.

The spectrum representing the unresolved nucleus was constructed from the 1-D nuclear spectrum that was extracted from the combined Gemini 2-D spectrum (Section~\ref{spectroastrometry}). However, as the 1 arcsec extraction aperture is $\sim 4\times$ larger than the HST PSF, the spectrum contains light from resolved (in the HST image) [O{\sc iii}] emission as well as  unresolved [O{\sc iii}] emission from the nucleus. To estimate and remove the contribution of the resolved [O{\sc iii}] emission, the [O{\sc iii}] and H$\beta$ narrow lines were fitted and subtracted from the nuclear spectrum, leaving only the continuum and broad emission lines (top and middle panels of Figure \ref{fig:offsetspectra_all}). Simulated spectra were then computed using the method described below, in which the unresolved nucleus was represented by the continuum plus broad line spectrum with varying fractions of the fitted [O{\sc iii}] and H$\beta$ narrow line spectrum added. This process was repeated until the intensity of the [O{\sc iii}] lines relative to that of broad H$\beta$ matched that of the original observed spectrum. In this way, we estimate that $90\%$ of the [O{\sc iii}] emission in the spectrum extraction aperture is unresolved in the HST image (i.e., it comes from the inner blue circle in Figure~\ref{fig:spec_rings}).

 We assume that the [O{\sc iii}] emission that is resolved in the HST image, but within 0.5$\arcsec$ of the nuclear point source, is dominated by the narrow line emission that is blended with the quasar nucleus emission within the Gemini extraction aperture. We represent this by the normalized [O{\sc iii}] and H$\beta$ spectrum obtained from the fit to the nuclear spectrum (middle panel in Figure \ref{fig:offsetspectra_all}). This was applied within the annulus between 0.12$\arcsec$ and 0.5$\arcsec$, shown by the blue and red circles in Figure~\ref{fig:spec_rings}), and represents the 1 arcsec extraction aperture from which the nuclear spectrum was obtained. This region encompasses most of the extended [O{\sc iii}] emission around the quasar. Finally, the [OIII] emission that extends beyond the central arcsecond, (the region beyond the red circle in Figure \ref{fig:spec_rings}) was represented by a spectrum extracted from a 1 arcsec aperture offset by 1 arcsec away from the nucleus (bottom panel in Figure \ref{fig:offsetspectra_all}).

 The resulting data cube was convolved with a PSF modeled by a Gaussian profile representing the typical seeing conditions for the Gemini observations and a simulated 2-D long-slit spectrum was extracted both in the North-South and East-West directions. Spectroastrometric measurements were then performed just as for the original spectrum by fitting Gaussians along the spatial direction at each wavelength. The resulting flux and displacement spectra are shown in Figure \ref{fig:spec2}. In the N-S direction The [O{\sc iii}]$\lambda 5007$ line shows a displacement relative to the continuum of $\sim 9$ mas to the North, while in the E-W direction, the displacement is $5.5$ mas to the East. 
 
The simulated displacement spectra therefore yield a displacement in the N-S direction which is in the same direction, but only 1/3 the amplitude of that measured directly from the Gemini data. In the E-W direction, the simulated displacement is actually in the opposite direction (East, whereas the measured displacement is to the West) and only $\sim 1/10 $ of the amplitude of the measured value. This result shows that the measured spectroastrometric displacement cannot be entirely explained by the asymmetric spatial distribution of the narrow line emission.

\subsection{Spectroastrometric continuum correction}
\label{contcorr}

As noted in Section~\ref{sec:specast}, the measured displacements will be smaller than the actual displacements due to dilution by the emission from the quasar itself (by which we mean the continuum and broad emission lines) and the underlying galaxy.
In order to correct the measured displacements for dilution and find the `true' displacement, we follow the analysis in \cite{porter04}, approximating the QSO and the host galaxy as point sources. Assuming that the host galaxy flux, $f_{\star}$, and the quasar flux, $f_{Q}$, vary slowly with wavelength ($\lambda$), then $f_{\star,j} \approx f_{\star}$ and $f_{Q,j} \approx f_{Q}$, where $j$ is the pixel index in the spectral direction. 
 We adapt eq. 4 in \cite{porter04} by adding a term representing the narrow line flux, $f_{NL,j}$, which is assumed to be spatially coincident with the point source representing the host galaxy. With this modification, 

\begin{equation}
\label{eq:dilution}
\begin{split}
\frac{\mu_{j}}{d} & = \frac{f_{NL,j} +f_{\star}}{f_{NL,j} +f_{\star} + f_{Q}} - \frac{f_{\star}}{f_{Q} +f_{\star}} 
\end{split}
\end{equation}

 \noindent where $\mu_{j}$ is the measured displacement, $f_{NL,j}$ is the narrow line flux, and $d$ is the actual displacement (in the same units as $\mu_{j}$).

We define 
\begin{equation}
\begin{split}
W_{NL,j} & = \frac{f_{NL,j}}{F_{C}} 
\end{split}
\end{equation}
\noindent and
\begin{equation}
\begin{split}
r_{\star} & = \frac{f_{\star}}{F_{C}}
\end{split}
\end{equation}

 \noindent where $F_{C} = f_{\star}+f_{Q}$ is the total continuum flux. By substitution in equation~\ref{eq:dilution} we get, 
  
 \begin{equation}
 \begin{split}
 \frac{\mu_{j}}{d} & = \frac{W_{NL,j}}{W_{NL,j}+1}(1 - r_{\star})
 \end{split}
 \end{equation}
 
 In E1821+643, the host galaxy light contributes only a small fraction of $F_C$ compared to the quasar nucleus,  hence 
 $r_{\star} < 0.1 $ (Section~\ref{Sec:hstdecomp}) and

\begin{equation}
\label{mu_d_eq}
\begin{split}
\mu_{j} & \approx \frac{W_{NL,j}}{W_{NL,j}+1 }d
\end{split}
\end{equation}

\begin{figure}
	\centering
    \includegraphics[width=\columnwidth]{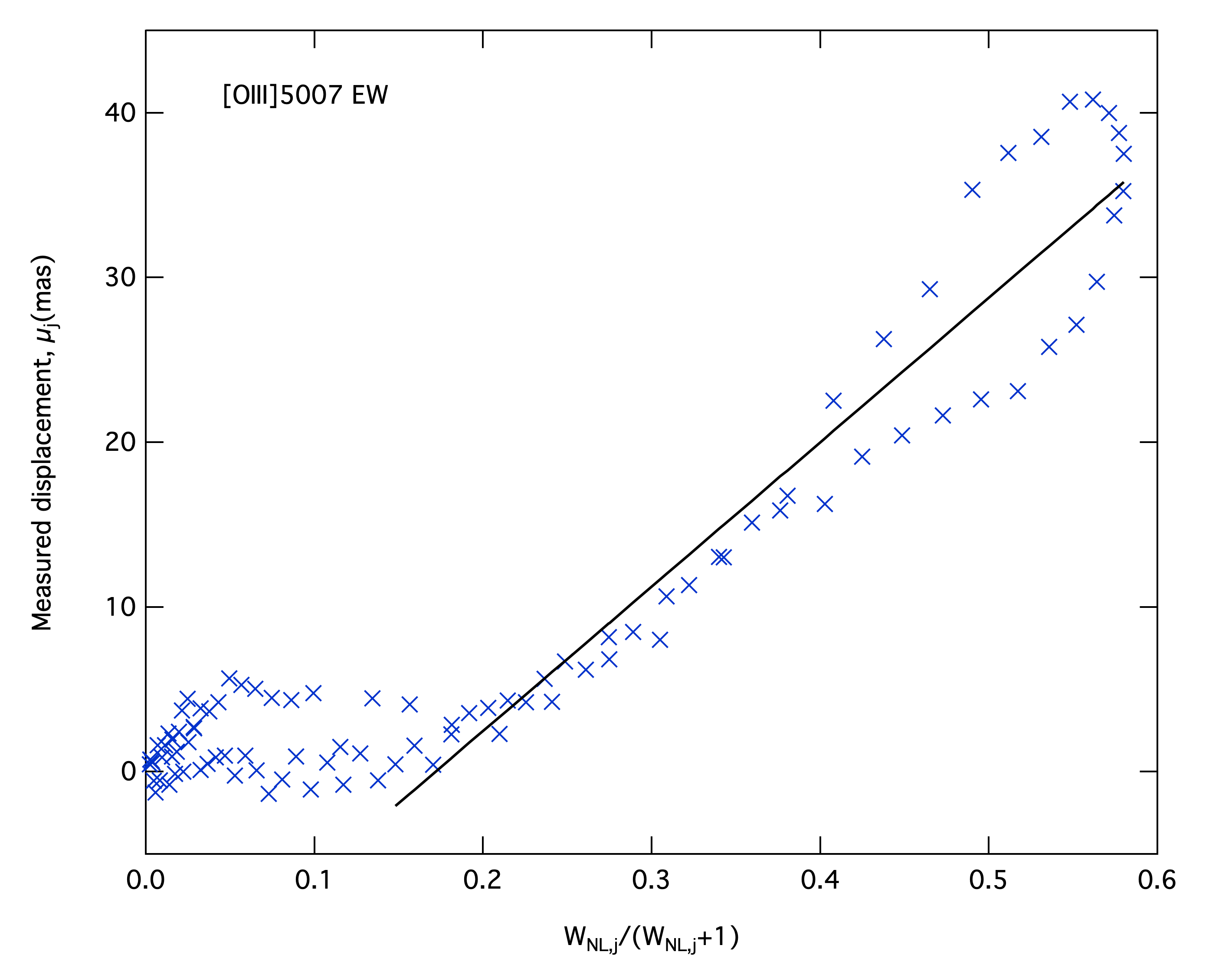}
	\caption{Spectroastrometric continuum correction. The relationship between the measured displacement and the ratio of the narrow line to continuum fluxes is shown for [OIII]5007\AA\ in the E-W direction. The slope of the relation yields the continuum-corrected displacement, $d$, (Equation~\ref{mu_d_eq}). The black line shows the linear fit to the data between approximately 6473\AA\ and 6504\AA, from which the value $d$ listed in (Table~\ref{mud}) was obtained. The ``flat'' portion of the relation around $\mu_{j}\sim 0$ is associated with the continuum on either side of the line. The `loop' at higher displacements arises from the asymmetry in the displacement profile (Figure~\ref{fig:dispo35007}, top left panel). }
	\label{fig:disp_corr}
\end{figure}

This model gives us a linear relationship between the measured displacement and the ratio of the narrow line to continuum fluxes, as given by eq.~\ref{mu_d_eq}.
 The slope, $d$, of this relationship is the corrected (''true'') displacement between the two sources. This was determined for both [O{\sc iii}] lines in the N-S and E-W directions from linear fits, with the values of $\mu_{j}$ and $W_{NL,j}$ taken from the corresponding displacement and flux spectra (Figure~\ref{fig:cont_fit}), respectively. An example is shown in Figure~\ref{fig:disp_corr}. The corrected values of the displacement for the two [O{\sc iii}] lines are given in Table~\ref{mud}, along with the measured values from Table~\ref{o3disp_table1} for comparison.  All values refer to the displacement measured at the center of the line. The quoted uncertainties on the $d$ values are those returned by the fit. The corrected displacements are consistent for both lines (as is expected) and have similar magnitudes of $\sim$80 - 90 mas, in both directions. If we assume the maximal estimated galaxy contribution, and set $r_{\star} = 0.1$, the corrected displacements increase relative to those listed in Table~\ref{mud} by approximately 10\%.

The corrected displacements given in Table~\ref{mud} correspond to a linear projected distance of $\sim$400pc in each direction and hence a radial displacement $\sim 580$\,pc to the NW, where the position angles are calculated from the corrected displacements. The results of the spectroastrometry analysis therefore indicate that the quasar nucleus -- i.e., the central continuum source and BLR -- is spatially offset by $\sim 580$\,pc to the NW of the gas producing the [O{\sc iii}] emission, i.e., the NLR. This seems consistent with the gravitational recoil hypothesis as proposed by R10, with the accretion disk and BLR remaining bound to the recoiling SMBH, whilst the NLR gas is left behind. In this case, the SMBH would be recoiling away from the galaxy center in the South - East direction.

\begin{table}
\centering
\caption{Measured and corrected spectroastrometric displacements for the [O{\sc iii}]  lines}
\label{mud}
\resizebox{\columnwidth}{!}{%
\begin{tabular}{lccccc}
\hline
\hline
Line & PA & $\mu_{NS}$ &$\mu_{EW}$ &$d_{NS}$ &$d_{EW}$ \\
     & $^{\circ}$ & mas & mas & mas & mas  \\
\hline

[OIII] 4959 \AA\, & -41.7 &14.9$\pm$1.0 &19.7$\pm$1.9 & 89.03 $\pm$ 6.72 & 79.22 $\pm$4.38 \\

[OIII] 5007 \AA\, & -44.1 &29.5$\pm$2.4 &39.2$\pm$1.2 & 90.64 $\pm$ 4.82 & 87.70 $\pm$ 3.87 \\
 
\hline

\end{tabular}
}

{$d_{NS}$ and $d_{EW}$ are derived from $\mu_{NS}$ and $\mu_{EW}$ respectively by using equation \ref{mu_d_eq} for the North-South and East-West direction.The PAs are calculated from the corrected displacement.}
\end{table}

\section{Discussion}

\label{e1821discussion}

R10 proposed the quasar E1821+643 as an SMBH gravitational recoil candidate based on  spectropolarimetric observations showing a velocity shift $\sim 2000$\,km\,s$^{-1}$ between the broad and narrow emission lines. If this interpretation is correct, one would expect the quasar nucleus to be spatially displaced from the centre of its host galaxy. R10 estimated that the SMBH could have moved a distance $\gtrsim 200$\,pc since the progenitor SMBH binary coalesced. This corresponds to an angular distance $\sim 40$\,mas, which is too small to be measured directly from HST images. The isophotal analysis technique employed by \cite{Batcheldor:2010aa} and \cite{Lena:2014aa} cannot be used since the quasar point source dominates the light from the host galaxy.

However, our spectroastrometry observations reveal spatial displacements in the [O{\sc iii}] lines of $\sim 80 - 90$ mas (after continuum correction) to both the North and West, relative to the optical continuum and broad emission lines (i.e., the quasar nucleus). The implied radial displacement of $\sim 130$\,mas corresponds to a projected distance $\sim$ 580\,pc in the North-West direction (Figure \ref{fig:cont_fit}). Given that the optical continuum is dominated by the quasar nucleus and that the NLR is located at the center of the host galaxy, this implies that the SMBH is spatially offset to the South-East relative to the galaxy center.

Our HST/ACS narrow band image resolves the [O{\sc iii}]  emission around the quasar and shows that it is asymmetrically distributed with respect to the unresolved nucleus. After subtracting the quasar point source, it can be seen that the [O{\sc iii}]  emission spans a wide arc in azimuth from the North-East around to the West and extends radially $\sim 0.5$\arcsec ($\sim 2.3$\,kpc) to the North-West of the nucleus. 

The quasar is slightly off-center to the South-East with respect to the surrounding [O{\sc iii}] emission, as would be expected if the measured spectroastrometric displacement represents a real spatial offset. However, the morphology of the [O{\sc iii}] emission is itself asymmetric, and this could also cause the spectropolarimetric displacement. 

We tested this possibility by constructing simulated spectroastrometry observations using the spatial distribution of the [O{\sc iii}] emission as mapped by the HST image.  
The displacements obtained in the simulations are much smaller in magnitude than the measured values (about 30\% and  $\lesssim 15$\% of the measured displacements along the North-South and East-West axes, respectively) and furthermore the displacement along the East-West axis is in the opposite direction to that observed (i.e., East rather than West). This indicates that the asymmetric structure of the [O{\sc iii}] emission can account for only a small fraction of the measured spectroastrometric displacement, which must therefore largely represent a real spatial offset.

At the inferred recoil velocity, the SMBH would retain its accretion disk and the bulk of the BLR, but the NLR (and the more extended [O{\sc iii}] emission) would be left behind R10. The observed spectroastrometric displacement spectrum can therefore be explained in terms of gravitational recoil of the SMBH: there is no displacement between the broad H$\alpha$ and H$\beta$ lines and the continuum, as the accretion disk and BLR remain bound to the SMBH, but the displacement in the [O{\sc iii}] lines can be attributed to the SMBH recoiling in the opposite direction (i.e., to the South-East). 
The size of the offset is a factor $\sim 3$ larger than the lower limit estimated by R10, but that is based on a jet advance speed of $0.1c$ and the actual speed could well be slower. R10 also argued that the recoil velocity vector should be roughly aligned with the radio axis, which is oriented NW-SE, consistent with the direction of the spectroastrometric displacement.

E1821+643 has been the subject of a long-term optical photometric and spectroscopic monitoring campaign \citep{2016ApJS..222...25S, 2017Ap&SS.362...31K, Kova2018}.
\citet{2016ApJS..222...25S} measured reverberation lags between the optical continuum and broad H$\beta$ and H$\gamma$ lines of 120\,d and 60\,d, respectively, and estimate that the SMBH has a mass $\sim 3\times 10^9$\,M$_{\odot}$. For this mass and the inferred recoil velocity, material orbiting within $R_{out} \sim 3$\,pc would remain bound to the recoiling SMBH (Equation~\ref{eq:rout}). Thus, $R_{out}$ easily encompasses the radius of the BLR implied by the reverberation lags. 

\citet{2016ApJS..222...25S} also found evidence for periodic variations in the optical continuum photometry with periods of $\sim 4000$, 1850, and 1200 days and in the spectroscopic data (broad H$\beta$ and H$\gamma$ lines and 5100\AA\  continuum) of $\sim 4500$ days. Further analysis by \citep{2017Ap&SS.362...31K, Kova2018}  yields broadly consistent results but also suggests that the same three periods ($\sim 12$, $\sim$ 6 and $\sim 5$ yrs) are present in both the continuum and emission line light curves. The origin of these periodicities is unclear. \citet{2016ApJS..222...25S} (hereafter referred to as S16) suggest that the longer periodic variations might be due to dense gas clouds orbiting a recoiling SMBH. On the other hand, \citet{2017Ap&SS.362...31K} argue that the longer periods may be associated with precession and orbital motion, respectively, in a binary SMBH, which they estimate would have a radius $\sim 6\times 10^{-2}$\,pc. This is $5\times$ smaller than the BLR radius as inferred from the H$\beta$ reverberation lag, implying that the BLR is circumbinary, rather than associated with one or both of the SMBH. This configuration does not offer a direct explanation for either the strongly asymmetric shapes of the broad line profiles, or the velocity shift relative to the narrow lines, or indeed the spatial displacement between the broad and narrow lines inferred from spectroastrometry. 

Alternatively, setting aside the question of the periodicities, the broad line velocity shift could conceivably be due to orbital motion in a binary SMBH, if the secondary is active and hosts a BLR. In this case, if the orbital angular momentum is inclined at an angle $i$ to the line of sight, the circular velocity would be $v_k \sim 2100/\sin i$\,km\,s$^{-1}$  and the binary separation would be $a \sim 3 \sin^2 i$\,pc (using the SMBH mass inferred by \citet{2016ApJS..222...25S}), corresponding to a period $P\sim 1.1\times 10^4 \sin^3 i$\,yr. We would not, therefore, expect to see significant changes in the broad line velocity shift over observable timescales. However, a ``hard binary'' such as this cannot explain the spectroastrometric displacement, and also seems inconsistent with the observed optical polarization (R10).

Overall, therefore, we consider that gravitational recoil of a post-merger SMBH provides a more complete explanation for the unusual properties of this object than a pre-merger binary SMBH.

A number of other recoil candidates have also been reported in the literature, identified either through velocity shifts between emission lines \citep[e.g.,][]{Shields:2009aa, Steinhardt:2012aa}, or spatial offsets \citep[e.g.,][]{Batcheldor:2010aa, Lena:2014aa}, but only a handful have been found that exhibit both. One of the best studied is CID$-$42 \citep{Blecha:2013aa, Novak:2015aa, Civano:2010aa, Civano:2012aa}
, a galaxy merger that contains two optical nuclei separated by $\sim2.5$ kpc, one of which is associated with an X-ray point source and a compact, flat spectrum radio source, indicating the presence of an AGN. The spectrum exhibits a weak broad H$\beta$ line which has a velocity offset $\sim 1360$\,km\,s$^{-1}$ relative to the narrow lines. In the radio galaxy 3C\,186 \cite{Chiaberge:2017aa}, the AGN is spatially offset by $\sim$ 11\,kpc relative to the center of the host galaxy and the broad lines have a velocity offset $\sim -2100$\,km\,s$^{-1}$ relative to the narrow lines. The quasar SDSS J0927+2943 exhibits a velocity shift of $\sim -2600$\,km\,s$^{-1}$ between the broad and narrow lines \citep{Komossa:2008aa} with the quasar nucleus being offset by $\sim 1$\,kpc from the centroid of the [O{\sc iii}] emission \citep{Vivek2009}. Another possible candidate is the triple merger system SDSS J1056+5516 \citep{Kalfountzou:2017aa}, which contains three emission-line nuclei, two of which have broad emission lines. The nuclei are separated by $15-18$\,kpc in projection, and are offset by $\sim 100$\,km\,s$^{-1}$ in velocity.  \cite{Kalfountzou:2017aa} suggest that, among other possibilities, one of the broad-line systems could be a recoiling SMBH.

Our results add E1821+643 to the short list of SMBH gravitational recoil candidates that exhibit evidence for both spatial displacements and large velocity offsets. Most of these, including E1821+643, have line of sight velocity shifts $\sim 1000$\,km\,s$^{-1}$, implying large kick velocities of several 1000\,km\,s$^{-1}$, perhaps in excess of the escape velocity of their host galaxies. Such large kicks are expected to be relatively rare events \citep{Lousto:2012aa} but tend to produce larger spatial offsets as marginally bound SMBH will spend most time near their large apocenters \citep{Blecha:2016aa}. In E1821+643, which has a relatively small spatial offset $(< 1)$\,kpc compared to several other recoil candidates, the SMBH must be in an early stage of its recoil, or approaching the pericenter on a return trajectory before the large amplitude initial oscillations become damped by dynamical friction \citep{Gualandris:2008aa}.

The displacement $\sim 100$\,mas that we measured in E1821+643 is too small to be detected by direct imaging with any existing telescope. It would only be marginally detectable even with the forthcoming James Webb Space Telescope (spatial resolution $\sim 0.1\arcsec)$. The offset in E1821+643 could, in principle, be detected with the proposed GRAVITY+ interferometer \citep{Eisenhauer2019} at the ESO Very Large Telescope (VLT), which will be capable of achieving milli-arcsec resolutions at magnitudes down to $K=22$. Unfortunately, E1821+643 is too far north to be observed with the VLT. It will be necessary to await the completion of the Thirty Meter Telescope before there is a Northern Hemisphere facility that has the capability to probe angular scales $\sim 10$\,mas, sufficient to confirm the spatial displacement of the SMBH in E1821+643.

The HST image also reveals a fainter region of extended [O{\sc iii}] emission in the form of a string of clumps spanning an arc from East to North, $\sim 3$\arcsec\ from the the nucleus (Figures \ref{fig:psfmanual} and \ref{fig:psfgalfit}). \cite{Aravena2011} detected CO emission centered $\sim 3$\arcsec\ South-East of the optical nucleus, which they suggest is a gas-rich companion galaxy that is merging with the giant elliptical hosting the quasar, or a remnant of a previous interaction. The arc-like [O{\sc iii}] structure may be part of a tidal tail, which is also a product of the interaction between the source of the CO emission and the quasar host. 

\section{Conclusions}

\label{e1821conclusions}

 Spectropolarimetry of the luminous quasar E1821+643 by R10 shows that the broad Balmer lines are redshifted with respect to the narrow lines and red-asymmetric in direct light but are conversely blueshifted and blue-asymmetric in polarized light. They propose that these unusual characteristics can be explained if the SMBH, with its retained accretion disk and BLR, is moving through the host galaxy at a speed $\sim 2000$\,km\,s$^{-1}$ due to gravitational recoil following coalescence of a progenitor binary SMBH. 

In this paper, we have used a combination of optical spectroastrometry and HST narrow-band imaging to look for the corresponding spatial offset between the quasar nucleus and the [O{\sc iii}] emission ``left behind'' by the recoiling SMBH. 

Our spectroastrometric analysis reveals spatial displacements in the [O{\sc iii}] lines of $\sim$80 - 90mas to both the North and West, relative to the optical continuum and broad lines. Assuming that the NLR is located at the center of the host galaxy, this is consistent with the quasar nucleus being offset by a projected distance $\sim 580$\,pc approximately to the South-East.

The HST/ACS images show that the nucleus is located within a region of extended [O{\sc iii}] emission which has a fan-like morphology spanning a wide arc from North-East to West. Another, clumpier, arc-like feature is seen in [O{\sc iii}] at $\sim 3$\arcsec in the North to North-East direction and may be a tidal tail resulting from a currently ongoing merger, or a previous interaction. 

The quasar appears to be located slightly to the South-East of the center of the main [O{\sc iii}] structure, consistent with the spectroastrometric displacement. The asymmetric morphology of the [O{\sc iii}] emission could itself produce a spectroastrometric displacement, but a simulated spectroastrometric observation based on the [O{\sc iii}] surface brightness distribution mapped by the HST narrow band image shows that it cannot account for the measured displacement in [O{\sc iii}], which we therefore attribute to a real spatial offset between the quasar nucleus and the surrounding [O{\sc iii}] emission region.

The combination of spectroastrometry and HST imaging thus provides evidence that the SMBH in E1821+643 is spatially offset from the center of its host galaxy by a projected distance $\sim 580$\,pc. E1821+643 can be added to the short list of SMBH gravitational recoil candidates that exhibit both large broad-line velocity shifts and spatial displacements.

\section*{Acknowledgements}
This research is based partly on observations made with the NASA/ESA Hubble Space Telescope obtained from the Space Telescope Science Institute, which is operated by the Association of Universities for Research in Astronomy, Inc., under NASA contract NAS 5-26555. These observations are associated with program G0-13385.

The work also made use of observations obtained at the international Gemini Observatory, a program of NSF's NOIRLab, which is managed by the Association of Universities for Research in Astronomy (AURA) under a cooperative agreement with the National Science Foundation. on behalf of the Gemini Observatory partnership: the National Science Foundation (United States), National Research Council (Canada), Agencia Nacional de Investigaci\'{o}n y Desarrollo (Chile), Ministerio de Ciencia, Tecnolog\'{i}a e Innovaci\'{o}n (Argentina), Minist\'{e}rio da Ci\^{e}ncia, Tecnologia, Inova\c{c}\~{o}es e Comunica\c{c}\~{o}es (Brazil), and Korea Astronomy and Space Science Institute (Republic of Korea).

\section{Data availability}
The HST data can be accessed from the Mikulski Archive for Space Telescopes (MAST) using the program ID 13385 at https://archive.stsci.edu/. The spectroscopy data can be accessed at the Gemini Observatory archive using the proposal ID GN-2010A-Q-103 at https://archive.gemini.edu/. The derived data generated in this research will be shared on reasonable request to the corresponding author.




\bibliographystyle{apj}

\bibliography{references_new}





\bsp	
\label{lastpage}
\end{document}